\documentclass{article}
\usepackage{epsfig}
\usepackage{amsmath}
\usepackage{amsfonts}
\usepackage{amssymb}
\usepackage{graphicx}
\usepackage{epstopdf}
\usepackage{amsmath}
\usepackage{graphicx} 
\usepackage{float}
\usepackage{amsthm,hyperref}

\begin{document}

\title{A Review of Methods for Estimating Algorithmic Complexity: Options, Challenges, and\\New Directions}
\author{
Hector Zenil\\hector.zenil@cs.ox.ac.uk\footnote{This review is based on an invited talk for MORCOM delivered at the International Society for Information Studies (IS4IS) summit at the University of California, Berkeley in June 2019.}\\Algorithmic Dynamics Lab, Karolinska Institute, Stockholm, Sweden;\\
Oxford Immune Algorithmics, Reading, UK; and\\
Algorithmic Nature Group, LABORES, Paris, France.
}
\date{}

\maketitle

\begin{abstract}
Some established and also novel techniques in the field of applications of algorithmic (Kolmogorov) complexity currently co-exist for the first time and are here reviewed, ranging from dominant ones such as statistical lossless compression to newer approaches that advance, complement and also pose new challenges and may exhibit their own limitations. Evidence suggesting that these different methods complement each other for different regimes is presented and despite their many challenges, some of these methods can be better motivated by and better grounded in the principles of algorithmic information theory. It will be explained how different approaches to algorithmic complexity can explore the relaxation of different necessary and sufficient conditions in their pursuit of numerical applicability, with some of these approaches entailing greater risks than others in exchange for greater relevance. We conclude with a discussion of possible directions that may or should be taken into consideration to advance the field and encourage methodological innovation, but more importantly, to contribute to scientific discovery. This paper also serves as a rebuttal of claims made in a previously published minireview by another author, and offers an alternative account.\\

\noindent \textsc{Keywords:} Algorithmic complexity; Kolmogorov complexity; practical feasibility; LZW; Shannon entropy; lossless compression; Coding theorem method; causality v. correlation; Block decomposition method; rebuttal to Paul Vit\'anyi's review.
\end{abstract}

\section{Introduction and preliminaries}

Researchers in the field of algorithmic complexity, also known as Kolmogorov complexity or Algorithmic Information Theory, are roughly divided between those interested in its most theoretical aspects, falling (for the most part) under the rubric of algorithmic randomness as applied to infinite sequences, and a much smaller group (discounting those who adopt the same popular statistical compression algorithms as the first group) interested in the methodological aspects of the theory of algorithmic complexity as applied to finite strings. For a review of the theoretical aspects of algorithmic randomness and progress made in the area the reader would do well to consult this good, up-to-date  survey ~\cite{franklin}, while this ~\cite{bienvenu3} provides an excellent historical account tracing how the field split into the above subgroups early in its history. While Kolmogorov was more interested in finite random objects~\cite{kolmo}, because `only finite objects can be relevant to our experience'~\cite{bienvenu3}, soon after its inception the field turned its attention to infinite sequences~\cite{martin} and the computability aspect of the question of `degrees of randomness', leaving only a handful of people following Kolmogorov's original path and driven by his primary motivation. This was chiefly on account of the restrictions imposed by the uncomputability of the measures of algorithmic complexity. Taken as its most salient feature it understandably led the larger group of researchers to interest themselves in its  most abstract---and restrictive---aspect.

In mathematics, restrictive results have a strong initial impact on seemingly promising research programs. For instance, it has been claimed~\cite{davis} that von Neumann decided to focus his research away from logic (at least for a time), as he thought G\"odel had delivered a decisive blow to research in the area. The mainstream answer to research in areas such as Set theory and First-order logic may have simply been `it is undecidable'. Nevertheless, decades of further research in logic would follow, which have resulted not only in theoretical advances, but in countless applications in wide-ranging domains.

In a similar (both theoretical and practical) fashion, restrictive results have been associated with the field of algorithmic complexity since its inception, but in contrast to previous restrictive results in other areas, theoretical advances~\cite{caludebook,nies,downey} have considerably overtaken methodological innovation in the area of applications~\cite{vitanyibook}. Turing's undecidability halting problem has been seen as an insurmountable obstacle to applications of algorithmic complexity. But again, many researchers did not give up, and looked for ways to circumvent theoretical limitations, setting their sights on the tremendous potential of applications of algorithmic complexity.

Here, I will briefly explain some challenges and limitations of lossless statistical compression algorithms in the study of the algorithmic complexity of finite strings. I then proceed to present evidence of how current methods for approximating algorithmic complexity do already produce considerably better results than popular statistical compression algorithms under certain regimes, while adhering more closely to the principles of algorithmic information and to the motivations of Kolmogorov, Solomonoff and Chaitin. Some of these motivations were the generation of knowledge, the improvement of scientific method, and a deeper understanding of the epistemological limits of humans and machines. 

Lempel-Ziv-Welch (LZW) is one of the most, if not the most popular statistical lossless compression algorithm based on dictionary methods for data compression~\cite{lzw}, and it underpins many algorithms and popular file formats. All variations of LZW (as used in popular file formats, and beyond) are based on the same statistical principle, and claims of `universality' refer to the set of assumptions under which the algorithm can count and group repetitions in a piece of data of arbitrary length (in principle), implementing and approaching optimal entropy rate.

Unless otherwise specified, I will use LZW to represent the set of statistical lossless compression algorithms based on classical information theory (as opposed to, e.g., methods such as algebraic compression~\cite{algebraic1,algebraic2}, or lossy compression), of which LZW is the most popular and widely used to allegedly approximate algorithmic (Kolmogorov) complexity. An overview of these compression algorithms is given in~\cite{datacompression}. LZW has been one of the most successful algorithms in computer science and has fulfilled its purpose as data compressor for files and images, as well as in offering an alternative order parameter that can legitimately be taken as a measure of `statistical complexity', sometimes called LZW-complexity. The argument in this review is that LZW-complexity, and its successes in its own right, should be clearly distinguished from algorithmic (Kolmogorov) complexity to allow the latter to make progress on the methodological front, especially because LZW cannot fully instantiate algorithmic complexity, while the part of it that it does instantiate can also be instantiated by classical information theory.

A simple statistical compression algorithm is called Run-length encoding (or RLE) and consists in encoding a string into consecutive segments of the form NX where N is the number of occurrences of X, and X is a digit that occurs N times consecutively in the original string. For example, $1112334$ would be `compressed' as $31122314$ which would read as 3 times 1, one time 2, two times 2 and one time 4. Evidently, the greater the number of consecutive digits the more effectively RLE compresses the original string. More sophisticated compression algorithms such as LZW, instead of counting the number of times that repetitions occur consecutively can have some memory (that allow building dictionaries) and also change their sliding window length generalising the behaviour of RLE on similar first principles.

A useful concept in this discussion (although not a central one) is also that of Borel normality. A Borel normal number~\cite{borel} has no repetitions and appears statistically random. Every segment of digits in the expansion of a Borel normal (or simply normal) number (in a given base) occurs with the same limit frequency. For example, if a number is normal in base 2, each of the digits `0' and `1' occurs half of the time at the limit; each of the blocks `00', `01', `10' and `11' occurs 1/4 of the time, and so on.

\subsection{Applications of Algorithmic (Kolmogorov) Complexity}

Cilibrasi, Li and Vit\'anyi were among the first to overcome the backlash against, as well as some of the theoretical obstacles standing in the way of applications based on or motivated by algorithmic complexity~\cite{cilibrasi,vitanyibook}. The use of lossless compression algorithms as a proxy for applications of algorithmic complexity opened the door to this innovation.

Nonetheless, such approaches have challenges and limitations of their own, and can only go so far. Theoretical and practical issues impede the use of statistical compression algorithms to the point of making them counterproductive for some general and many specific applications. Naturally, in view of this fact, other researchers are exploring different avenues, in pursuit of novel approaches and methodological innovations. I contend that complementary numerical approaches to algorithmic complexity offer a promising direction for the area that will feed back and forth between theory to applications.

Central to algorithmic information is the concept of algorithmic complexity. Briefly, the (plain) algorithmic complexity~\cite{kolmo,chaitin} of a string $s$ (also known as Kolmogorov, or Kolmogorov-Chaitin complexity) denoted by $C$, is defined as the length of the shortest computer program that computes (outputs) $s$ and halts, where the program runs on a given reference universal computer. The invariance theorem (and conditions under which it holds), in-depth details of which are provided  in~\cite{vitanyibook}, posits that there exist some reference universal machines (called `optimal') for which the invariance theorem holds, establishing that the difference between estimations using different Turing-complete languages is bounded by a constant independent of the string $s$. Formally, this means that under the assumption of reference machine optimality, there is a constant $c$ for all strings such that $|C_{U'}(s) - C_{U''}(s)|\leq c$. A non-optimal Turing machine $U$ can be illustrated by constructing a Turing machine such as the one given in~\cite{review} that applies an operation resulting in estimated complexity values as a partial function of the string. For example, a machine that multiplies a short description of $s$ by $n$, where $n$ is the number of non-zero digits found in the input string. Clearly, this machine approximates algorithmic complexity as a function of a property of the string rather than independently of it, and the difference cannot be bounded by a constant $c$ for all strings, so the invariance theorem does not hold. When this is the case, not even (relative) complexity ranking invariance is achieved. If, however, the invariance theorem holds, nothing guarantees quick convergence. Non-optimal machines, however, are constructed in a rather artificial fashion, and to the author's knowledge the conditions under which non-optimality results from natural choices of computing model remain an open question.

When statistical lossless compression algorithms---such as those based upon LZW and cognates---were adopted as a tool to estimate algorithmic complexity, the apparently successful results gave applications credibility among some practitioners (though not so much among theoreticians). The lossless compression approach to algorithmic complexity worked for reasons not directly related to the theory of algorithmic complexity (or as a result of it), but because of the already established connection with Shannon entropy. The Shannon entropy $H$ of a given discrete random variable $s$ with possible values $s_1, \dots s_n$ and probability distribution $P(s)$ is defined as $H(s)=-\sum_{i=1}^n P(s_i) \log_2 P(s_i)$ (if $P(s_i) = 0$ for a given $i$, the value of the corresponding term is 0). Roughly, the relationship between $H(s)$ and $C(s)$ for a string modelled as a random process~\cite{zenilentropy} is that the expected values of $H(s)$ and $C(s)$ converge up to a constant for computable distributions ~\cite{antunes}. This is true if we have full access to the ground-truth probability distribution for $H(s)$, and for $C(s)$ we assumed the decidability of the halting problem. However, when the nature of the source of an object is unknown and is modelled~\cite{zenilentropy} as a random variable, if $H(s)$ indicates non-randomness by assigning a small value, then $C(s)$ is not algorithmically random, but when $H(s)$ suggests randomness $C(s)$ does not necessarily follow. In contrast, if $C(s)$ is algorithmically random, $H$ cannot be statistically non-random. It follows that $C$ is then a generalisation of $H$ and requires fewer assumptions. In other words, $C$ offers a means of inferring the underlying probability distribution. Like $H$, algorithms such as LZW cannot characterise algorithmic randomness because they implement a version or function of Shannon entropy (entropy rate), not only because they are not universal in the Turing sense, but also because they are designed only to characterise statistical redundancy (repetitions). However, accounting for non-statistical regularities ought to be crucial in the theory of algorithmic information, since these regularities represent the chief advantage of using algorithmic complexity.

\subsection{Applications of Algorithmic Probability}

Algorithmic probability~\cite{solo,solo2,solo3} (AP) and the Universal Distribution~\cite{levin} approach the challenge of algorithmic inference from the standpoint of the theory of computation. Vit\'anyi and colleagues have given a wonderful account of the properties of this mathematical concept~\cite{miracle}. Formally, algorithmic probability~\cite{solo,levin} is defined as $AP(s) = \sum_{p:U(p) = s} 1/2^{|p|}$, where $p$ is a random computer program in binary (whose bits are chosen at random) running on a so-called prefix-free (with no program being a proper prefix of any other program) universal Turing machine $U$ that outputs $s$ and halts. 

Both algorithmic complexity and algorithmic probability are not computable; they are semi-computable (upper and lower), meaning that approximations from above and below are possible and are deeply related to each other. A formal connection is established by the so-called (algorithmic) coding theorem~\cite{vitanyibook} that establishes that a short computer program is also algorithmically highly probable and vice-versa. Formally, the coding theorem~\cite{levin} states that $C(s)= -\log AP(s)+O(1)$. AP and the coding theorem hold for prefix-free complexity, but given that the difference with respect to $C(s)$ is a slightly larger  $O(log^2 |s|)$ term~\cite{solovay}, we will not make further distinctions between plain and prefix-free algorithmic complexity but it is worth to be aware of the subtleties~\cite{antunes2}.

The incomputability of $C$ and $AP$ has meant that for decades after the discovery of these measures very few attempts were made to apply them to other areas of science, with most researchers' reaction to the field, especially as regards applications, being sceptical, their reservations based on claims of incomputability. Nonetheless, in a very fundamental way, algorithmic probability (and hence algorithmic complexity) can be regarded as the ultimate theory for scientific inference~\cite{miracle,minsky}, and we took this to heart when attempting, regardless of the attendant challenges, to find measures related to AP~\cite{algonat,d4,plos}. 

The original formulation of algorithmic probability is of fundamental interest to science because it can address some of its most pressing challenges, such as inference, inverse problems, model generation, and causation, which happen to be the topics of interest in our research programme (beyond simple estimations of algorithmic complexity represented by a single real-value number). This relevance to science has been touted by Vit\'anyi himself, and colleagues, in a very engaging article~\cite{miracle}, and has more recently been pointed out in areas such as AI and machine learning by people such as Marvin Minsky, who claimed that Algorithmic Probability was one of the most, if not the most important theory for applications~\cite{minsky}. Approaches to algorithmic complexity and algorithmic probability that take into consideration finite resources have been proposed before, such as resource-bounded Kolmogorov complexity~\cite{levin,levinkt,buhrman,allender} and Universal Search for applications~\cite{levin,schmid2,hutter} based on, for example, dovetailing all possible programs and cutting short runtimes. A good introduction may be found in~\cite{vitanyibook}.

For example, motivated by AP and often said to be based on the concept of Occam's razor, but designed to circumvent the incomputability of AP, some methods such as Minimum message length (MML)~\cite{wallace} and Minimum description length (MDL)~\cite{rissanen} were introduced as methods for statistical inference that avoid universality and therefore incomputability by assuming that Occam's razor or some part of AP could be instantiated by Bayesian or traditional statistical learning from data. In the case of MML, it is a Bayesian approach that attempts to represent beliefs about the data-generating process as a prior distribution and separate data from model in a similar statistical fashion to  MDL, on which it is largely based.

The notion behind AP is very intuitive. If one wished to produce a segment of the digits of $\pi$ randomly, one would have to try time after time until one managed to hit upon the first numbers corresponding to the segment of an expansion of $\pi$, which is strongly believed to be absolutely Borel normal~\cite{borel}. The probability of success is extremely small: $1/10$ digits multiplied by the desired quantity of digits. For example, $1/10^{2400}$ for a segment of 2400 digits of $\pi$. But if instead of shooting out random numbers one were to shoot out computer programs to be run on a digital computer, the result would be very different.

A program that produces the digits of the mathematical constant $\pi$ would have a higher probability of being produced by a computer. Concise and known formulas for $\pi$ could be implemented as short computer programs that would generate any arbitrary number of digits of $\pi$, but to find these generating formulas one would need something other than a purely statistical method; one would need a mechanism able to produce Turing-complete computer programs in the first place.

Despite the simplicity of algorithmic probability and its remarkable theoretical properties, a potentially large constant slowdown factor has kept it from being much used in practice. Some of the approaches to speeding things up have included the introduction of bias and making the search domain specific. It was not until we proposed numerical estimations based upon AP~\cite{d4,plos,bdm} that applications of AP to diverse areas of science began to be conceived and explored, including our own most recent applications~\cite{nmi,iscience,nmi}. We see our work as having offered an opportunity to open up a discussion and to explore alternatives to the otherwise dominant use of statistical lossless compression (see Table~\ref{regimes2}). In our approach we have explored the effect of relaxing, in a higher, finer-grained, recursive fashion, some of the conditions of weakening Turing-universality~\cite{liliana} (as opposed to embracing a weak computational power needed to instantiate
statistical compression). What we have demonstrated is that AP may account for up to 60\% of the simplicity bias in the output distribution of strings at each level of the Chomsky hierarchy~\cite{chomsky} when running on random programs. There are practical applications of AP that make it very relevant. If one could translate some of the power of algorithmic probability to decidable models (below Type-0 in the Chomsky hierarchy~\cite{chomsky}) without having to deal with the incomputability of algorithmic complexity and algorithmic probability, it would be effectively possible to trade computing power for predictive power. Other interesting approaches can be identified with the field of computational mechanics~\cite{crutchfield}, where a combination of statistical approaches to constructing finite state models has been proposed.

A more comprehensive review than the one offered in the short review in~\cite{review} would have covered statistical lossless compression as the dominant technique in the area of applications based upon or motivated by algorithmic complexity. While authors like Li, Cilibrasi, Vit\'anyi and others helped advance the potential practical side of a previously heavily theoretical field, the heuristics used have also contributed to some methodological stagnation, and to the predominance of a broad and wild programme that uses (and often abuses) heuristics based on statistical lossless compression algorithms such as LZW, (largely) unrelated to algorithmic complexity in any meaningful way (beyond, e.g., Shannon entropy rate).

In the past, researchers have however explored properties that conform with the theory of algorithmic information but are relaxed for applications. The one that is most commonly relaxed is Turing universality, as in approaches like MDL, MML, LZW and cognates, which thus never face any other challenge relative to the hierarchy of features that pertain to the theoretical specification (see Table~\ref{relax}). 

In many such approaches the basic idea is to find the hypothesis that statistically compresses the data best. But methods of very different sorts can be applied, some of them more relevant to algorithmic complexity than others. Traditionally, the model associated with the data compressor is not entailed in the method itself but is proposed or found by the researcher, and often pre-selected based on a feature of interest (e.g. repetitions for LZW). The inference model is missing and cannot be fully explored because these approaches give up on Turing-completeness, denying themselves the key computational element of algorithmic probability, which may provide such a model generation framework.

Other approaches, less often used, may adopt Turing-universality, but are then forced to assume or constrain their exploration by, e.g., limiting resources, such as in resource-bounded complexity and resource-bounded algorithmic probability~\cite{antunes2}; and still others have exchanged computability for model stochasticity, as in some approaches from computational mechanics. Some approaches remain pure, like certain approaches to program synthesis and automatic theorem proving, making them methodologically more difficult to apply because of the resources they need, though they are also more relevant in the context of algorithmic probability.

\subsection{Beyond incomputability}

Ultimately, what Shannon entropy is meant to quantify is statistical typicality, that is, how common or rare a distribution of elements of an object appears to be relative to the rest of the objects. The greater the Shannon entropy, say in bits, the more random looking. In contrast, what algorithmic complexity is meant to quantify is algorithmic randomness. When an object cannot be compressed beyond its uncompressed length, the object is algorithmically random. Algorithmic complexity was conceived to answer the question of whether statistical randomness was the only kind of randomness and whether statistical randomness could characterise all types of randomness, understood in the broadest sense. Indeed, algorithmic complexity is one of those concepts for which, at least at a basic level, different formulations based on uncompressibility, unpredictability, and atypicality, converge~\cite{levin,schnorr,schnorr2,martin}, making it formally the agreed-upon quintessential definition of randomness in mathematics (and thus in science). Algorithmic complexity answers in the negative, with very simple examples, the question about whether statistical randomness can or cannot capture the intuitive notion of randomness in this broader sense. It does not even come close, because it misses a whole set of non-random objects that can be recursively compressed, that can be predicted, and that are atypical in the sense that they do not have the features that most other objects have. These features, when present, allow compression, but when absent do not (or prediction, for that matter~\cite{schnorr}). It is therefore key for algorithmic complexity to be able to distinguish, or have the capacity to potentially distinguish, aspects of algorithmic randomness from statistical randomness.

For this reason, the field of algorithmic information has the potential to transform science and scientific methodology, to raise the bar for scientific evidence and scientific proof, through control experiments that rule out stronger, less fragile kinds of randomness as well as
the kinds of spurious correlations pervasive in science~\cite{calude1,smalldata} inherited from purely statistical approaches (as opposed to causal and state-space descriptions).

We believe this can be achieved by moving from the currently dominant data-reliant scientific methods to model-centred ones. Unfortunately, the area of algorithmic information has had little impact to date, and has often taken the wrong direction when applied to science, continuing to use what ought to be superseded, yet all the while giving the contrary impression. Indeed, we have focused for far too long on a single approach based on statistical lossless compression algorithms, and methodological innovation has come at the cost of scepticism from established researchers, even those who once promoted innovation.

\section{Alternatives to lossless compression}
\label{ctm}

In~\cite{d4,plos}, a measure motivated by algorithmic probability was introduced that entails exploring an enumeration of Turing machines running on a reference Universal machine that is assumed to be optimal and whose variant specifications across several papers are widely used in other areas of the theory of computation, a machine based on Rado's so-called Busy Beaver problem~\cite{rado}.

Let $(n,m)$ be the space of all $n$-state $m$-symbol Turing machines, $n,m > 1$ and $s$ a sequence. Then we define CTM (standing for Coding Theorem Method) as: $CTM(n,m)(s) = \frac{|\{T \in (n,m): T \textit{ produces $s$}\}|}{|\{T \in (n,m)\}|}$.

What CTM does is to precompute these programs for a large set of short strings and combine them with classical information theory using what we call BDM (standing for Block Decomposition method) in order to build a short program based on small pieces than can explain a larger piece of data for which CTM has not (yet) been calculated. CTM can also be seen as exploring all possible computable compression algorithms, each represented by a decidable (time-bounded) Turing machine from a proper universal enumeration, whose machine number is stored and allows for a decompression process, thus honouring the true spirit of Kolmogorov complexity as the ultimate data compression method, rather than privileging its much more shallow connection to statistical compression only.

CTM is motivated by the relation~\cite{plos} between the algorithmic probability of an object such as a string (the frequency of production of a string from a random program) and the algorithmic complexity of the said string (the length of the shortest description under the reference machine). 

Just as resource-bounded complexity compromises on resources like runtime, and MDL, MML and LZW give up on Turing completeness (and therefore on algorithmic complexity for the most part) in order to model data, and LZW is intended to capture no other feature of an algorithmic nature beyond repetitions in data, the assumptions that CTM makes are that the invariance theorem is valid and that the additive (or other) constant is small enough not to impact the ranking or the top ranking of the smallest computable models explaining data, which is precisely what CTM investigates empirically with a view to making changes to the underlying computing models and testing against the data. 

CTM is intended not just as a theoretical construct, but as a technique to be deployed in practice. CTM outperforms or complements lossless compression algorithms and other techniques such as MDL in various respects. For example, CTM can provide a finer-grained classification of short strings and is able to provide computable descriptions in the form of specific computer programs in the language of Turing machines, rather than black boxes (compressed files) with little state-to-state correspondence, and most likely unrelated to the generating process other than in terms of trivial statistical redundancy. MDL, MML and other such approaches present their `flaws' as features (or even advantages), such as the claim that MML does not require Turing completeness or that MDL is computable (without ever mentioning what's sacrificed). MDL avoids assumptions about the data-generating process. 

One way to see CTM is as effectively exploring the space of all the possible computer programs able to compress a piece of data. This is in contrast to settling on a single program that one can ascertain in advance to be incapable of dealing with any algorithmic feature in data (as opposed to just statistical features, i.e. substring repetitions).

The Block decomposition method (BDM) is an extension of CTM defined as $BDM(s)=\sum_i CTM(s_i) + \log (n_i),$ where each $s_i$ corresponds to a substring of $s$ for which its CTM value is known and $n_i$ is the number of times the string $s_i$ appears in $s$. A thorough discussion of BDM may be found in~\cite{bdm}. BDM extends the power of CTM to deal with larger strings by sticking together, using classical information theory, the estimations of complexity values of shorter substrings based on CTM under the universal model of computation.  BDM is explored in~\cite{bdm}, and among its properties is the fact that it cannot perform worse than Shannon entropy but can indeed improve upon it when combining values of algorithmic complexity for its substrings.

BDM glues together the programs producing each of the data pieces using classical information. The result is a hybrid measure that provides a candidate upper bound of algorithmic complexity.

In contrast to statistical compression algorithms, CTM can, in principle, deal with simple objects such as $123456\ldots$ in a natural way (or $1,2,3,4,5,6,\ldots$ seen as a data stream from a generating process) by identifying the underlying successor function in any base from observing the piece of data and extending the Turing machine rule space, as shown with 2 examples in~\cite{nmi}. And even for $\pi$~\cite{bdm}, both in principle and in practice, as suggested in~\cite{bdm}, when normalising by highest values Shannon entropy (and therefore LZW and cognates) will by definition  retrieve maximal randomness, but for CTM and BDM, it does not have maximal randomness and actually shows a trend of decreasing values. What CTM does is find the set of computer programs that can explain segments of $\pi$, something that statistical compression cannot do.

\begin{figure}[ht!]
\centering
\scalebox{.3}{\includegraphics{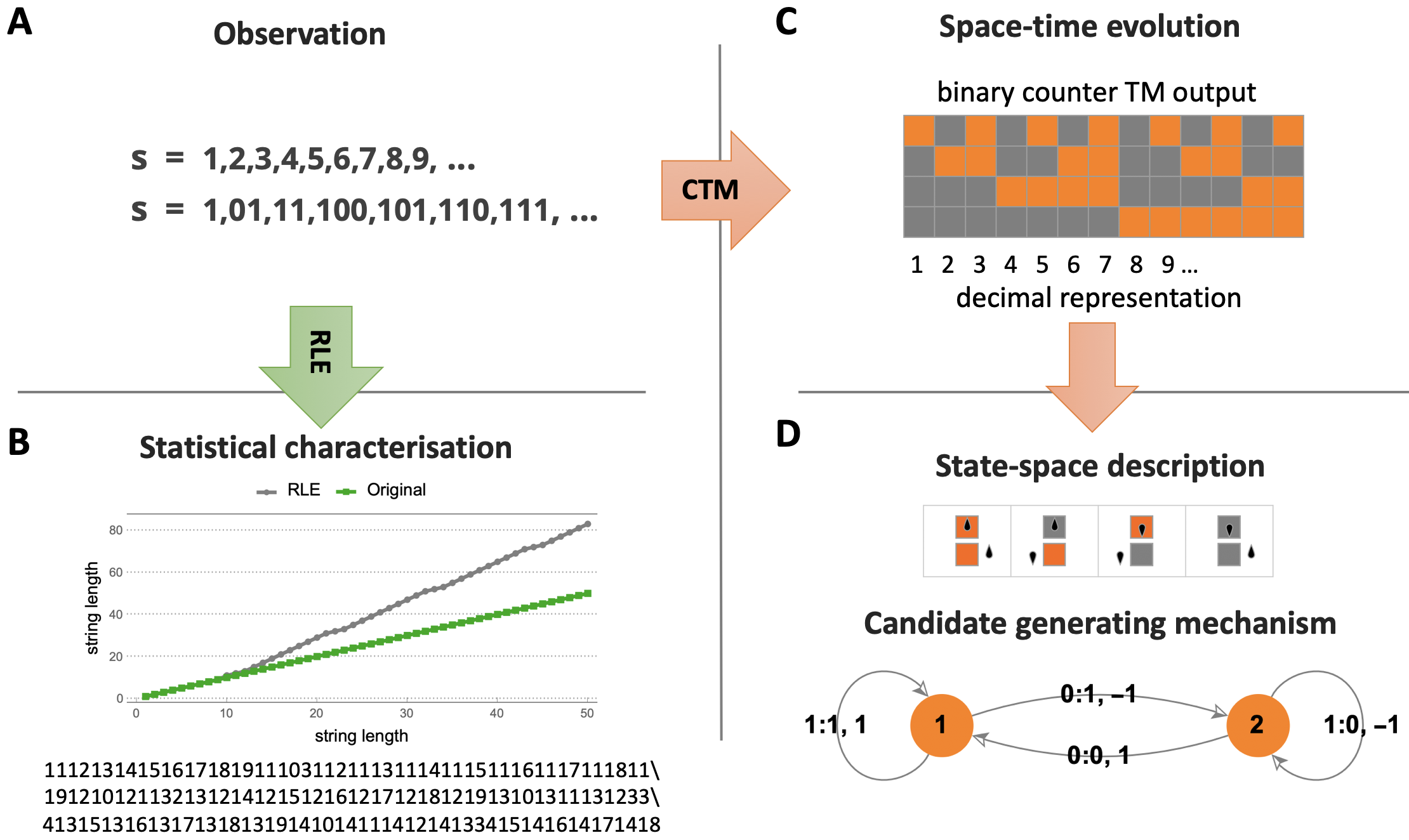}}
\caption{\label{program}(A) An observer trying to characterise an observed stream of data ($s$) of unknown source (in this case the sequence of natural numbers in decimal or binary that the observer may not recognise) has currently two options: (B) the statistical compression approach (green arrow) represented by Run-length encoding (RLE) producing a rather cryptic and even longer description than the original observation with no clear correspondence between possible ground and represented states; or, alternatively, (C) an approach that takes on the challenge as in an inverse problem represented by CTM (green arrow) that allows to find the set of small computer programs according to some reference machine up to certain size that match their output to the original observation thereby potentially reverse engineering the generating mechanism that may have recursively produced the sequence in the first place. Such an approach allows a state-space description represented by a generative rule or transition table and a state diagram whose elements may correspond (or not) to a physical unfolding phenomenon against which it can be tested (notice that the binary to decimal TM is also of finite small size and is only an intermediate step to arrive to $s$ but can be one of the engines behind, here for illustration purposes only the TM binary counter is shown as an example).}
\end{figure}

CTM (and BDM) offers an alternative to statistical compression algorithms (see Table~\ref{regimes2}) but extends beyond an exclusive interest in algorithmic complexity into the area of causal discovery (see Fig.~\ref{program} and Fig.~\ref{causalgaps}); it is an alternative that not only in its application but also in its approach is radically different from popular statistical lossless compression algorithms like LZW. The difference is that we already know that statistical compressors are at variance with the theory and unable to produce a state-space hypothesis. In contrast, CTM is able to find computer programs that offer a state space analysis capable of producing generative mechanisms of the observed data that may better fit the state space of the actual phenomenon producing the data (see~\ref{program} for an example).

This is a list of some of the suggestions that CTM and BDM do not make or prove: that algorithmic complexity is computable; that the reference universal machine is optimal (or that it is not); that the assumed additive constant is under control, or even that the invariance theorem applies (c.f. optimality, previous point); or that the complexity ranking and estimations are robust or invariant (despite statistical evidence indicating otherwise). On the other hand, this is what CTM and BDM do offer: a better-grounded alternative to the current but---for most purposes---mostly incorrect approaches to algorithmic complexity (when claimed to be related to algorithmic complexity) consisting of using only statistical lossless compression algorithms such as LZW when no full disclosure of its divergence from algorithmic complexity is made; it opens up an empirical approach to testing questions related to necessary conditions in real applications, including the questions of machine optimality and additive constants, robustness and convergence (or divergence) of output distribution (related to the invariance theorem and the Universal Distribution itself).

Approaches that aim to apply algorithmic probability have been advanced in the past, and are interesting. One of these consists of a so-called Universal Search~\cite{levin} based on dovetailing all possible programs and their runtimes, such that the fraction of time allocated to a program is proportional to program size (in number of bits). Despite the algorithm's simplicity and remarkable theoretical properties, a potentially huge constant slowdown factor has kept it from being much used in practice, or at any rate used more frequently than compression algorithms.

Some of the approaches to speeding it up have included the introduction of bias and making the search domain specific, which has at the same time limited the power of algorithmic probability (AP). We settled on a measure that seemed deeply in harmony with the spirit and definition of algorithmic probability and Levin's Universal distribution, but which has never been presented as being exactly one or the other. 

There are also deeper and practical applications of measures directly related to AP, whether based on or motivated by it, that make them relevant. If one could translate some of the power of algorithmic probability to, for example, decidable models, and study them without having to deal with the incomputability of algorithmic complexity and algorithmic probability, and without resorting to weak methods such as statistical compression algorithms, but retaining some of the computational power to characterise recursive randomness, it would be effectively possible to trade computing power for predictive power. This is something we have investigated by relaxing the assumptions even further, beyond Turing universality, by constraining the computational power of computational models~\cite{liliana}. The relationship and a smooth trade-off found before reaching the undecidability frontier mean that algorithmic complexity (and algorithmic probability) are also relevant, and that some theory can be retained and partially recouped from simpler models of computation in exchange for computational power.

\subsection{Time to stop fooling ourselves}

Researchers interested in applying algorithmic complexity to their data for purposes of, e.g., model or feature selection, classification, clustering or data dimension reduction, had traditionally been forced, indeed had no other option but to use statistical lossless compression algorithms. During the past few years, however, CTM has offered a refreshing alternative that is methodologically very different from compression. It is complementary in that it is better in a regime where compression is well known to fail (small objects) and, in the long run, when combined with BDM, it has proven to behave just like a statistical lossless compression algorithm in the worst case~\cite{bdm}, with the advantage that CTM, which is at the core of BDM, can potentially offer local approximations to algorithmic complexity by finding short computer programs able to explain/generate a small piece of data. CTM and BDM can thus constitute a true proxy for and generalisation of statistical compression algorithms.

In speaking of the obfuscation of the differences in approaches and of how statistical compression has been used on weak grounds before, let us cite the same mini review~\cite{review} when it claims that `[We] can approximate the Kolmogorov complexities involved by a real-world compressor. Since the Kolmogorov complexity is incomputable, in the approximation we never know how close we are to it.' This is a known fact in the application of data compressors but unfortunately is not disclosed as often as it should be in papers using compressors to allegedly estimate algorithmic complexity.

The short review in~\cite{review} closes one of its sections with a defence of past heuristic approaches based on Shannon entropy, a defence based on an argument about `natural data' and on the concept of a `construct' as something artificial or unnatural and unlikely to be found in the real world: `In fact, we assume that the natural data we are dealing with contains primarily effective regularities that a good compressor finds.' In transiting to the next sentence a huge leap is made: `Under those assumptions the Kolmogorov complexity of the object is not much smaller than the length of the compressed version of the object.'

It is time to stop believing that we are doing something more interesting than measuring basically Shannon entropy and counting repetitions in data when using lossless compression algorithms such as LZW.

The statistical compression approach to algorithmic complexity appears to give up on algorithmic complexity itself by relaxing what makes it possible, at the very first level of challenges in the hierarchical table of requirements and conditions in Table~\ref{relax}, relaxing the most basic of the requirements, that of Turing-completeness (level 0) and embracing only statistical redundancy, which Shannon entropy alone can deal with (as it does with LZW), thus ignoring the problems posed by deeper theoretical challenges, including those of incomputability and optimality.

\subsection{The regime of short strings}

The short review in~\cite{review} also claims that one or more of the papers about CTM mistakenly referred to `the shortest computer program’ when referring to cases in which, according to a well-specified enumeration, CTM was able to find the shortest Turing machine starting from smallest to largest. According to the review, the phrasing in~\cite{plos} should have been `a [sic] shortest computer program'.  This is a very weak and unfounded criticism, as statements to `the shortest computer program’ were only made after introducing in full detail the specification of the reference universal machine, making the charge irrelevant.

Another criticism of CTM in~\cite{review} was that it is only able to estimate values for strings shorter than 20 or 32 bits. Not only do other researchers see this as its most important feature~\cite{personal}, but CTM was first and foremost designed to cover the regime of short strings, as no other method was available, not even data compressors, to deal short strings. Thus in being able to complement lossless compression approaches to algorithmic complexity it immediately made a contribution, providing  objective estimates, even though based on assumptions of optimality and fast convergence. CTM offered the first ever classification and ranking of short strings (see Fig.~\ref{models}). This regime of short strings is of no small interest; they are among the most relevant for real-world applications such as perturbation analysis~\cite{iscience}, molecular biology and genetics~\cite{nar}.

The brief review in~\cite{review} fell short when disclosing the challenges and limitations that the only other currently available option purportedly able to estimate algorithmic complexity faces. Indeed, statistical lossless compression algorithms such as LZW, which are unable to deal effectively with strings with fewer than 1000 bits, collapse string values (see Fig.~\ref{diversity}), offering little understanding of a large set of strings, as compression algorithms introduce variable length instructions that effectively contribute to the additive constant, resulting in a constant equivalent to that which CTM must also deal with, and where check-sum codes contribute to a steep step-like behaviour that makes values irrelevant for most purposes related to algorithmic complexity, though legitimate when presented as a different order parameter such as LZW-complexity as long as the distinction is made and no further unjustified connection to algorithmic complexity is claimed.

\begin{figure}[ht!]
\centering
\scalebox{.23}{\includegraphics{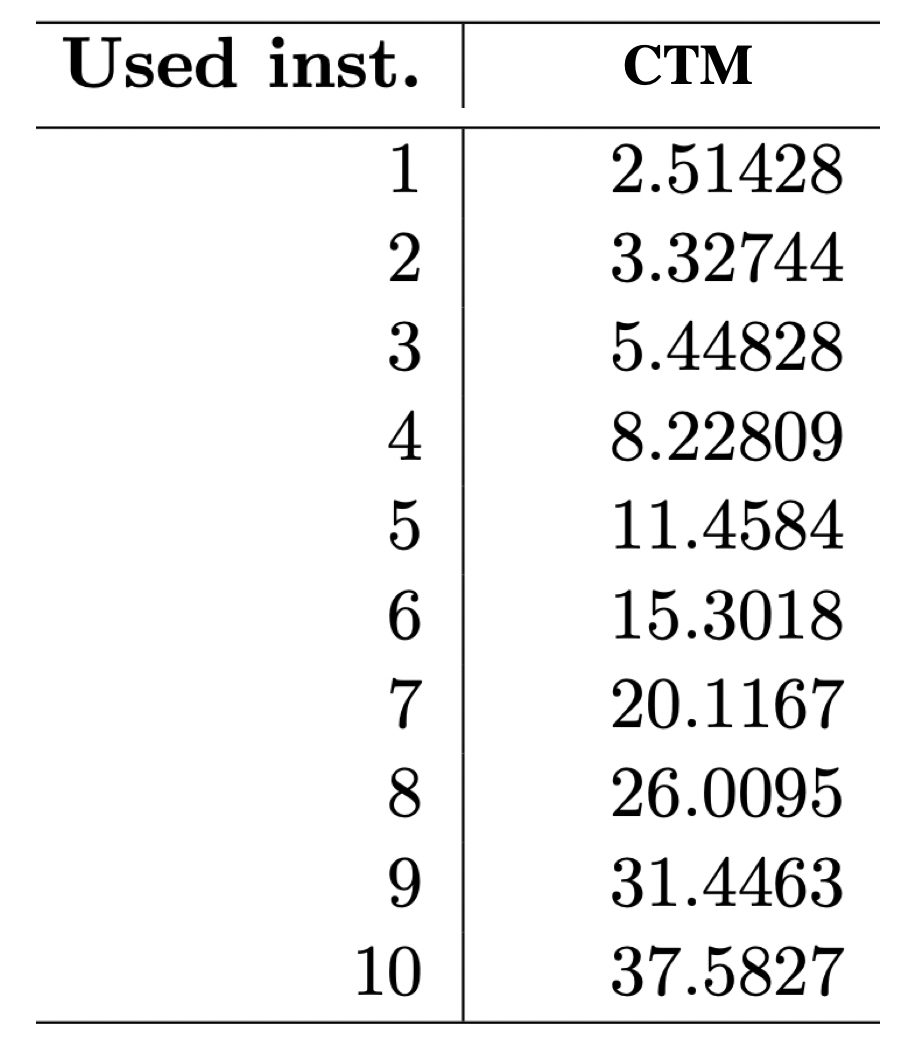}}
\caption{\label{instructions}Test for model internal consistency and optimality. The number of rules  used in the transition table for the first and shortest Turing machine producing a string following the enumeration is in agreement with the estimation arrived at by CTM after application of the coding theorem under assumptions of optimality. Adapted from~\cite{computability}}
\end{figure}

Even assuming that empirical distributions built from CTM are significantly affected by factors such as the (non-)additive constant (wilder if optimality does not hold), they do produce testable outcomes in the form not only of real-value numbers as approximations to algorithmic complexity, but also as a set of finite state computer programs able to explain the data under analysis. Indeed, outcomes can be tested even against compression algorithms, because what is highly compressed should have low algorithmic complexity according to CTM (and BDM), and what is not compressible should also tend to have higher CTM (and BDM), which is in fact what happens~\cite{bdm}. CTM has passed dozens of tests (e.g.~\cite{computability,bdm,physicaa,kolmo2d}), conforming both with theory and intuition, some of which included, e.g. tests of method internal self-consistency, such as comparing the estimations of algorithmic complexity after the application of the coding theorem, on the one hand, and the assumed approximation to algorithmic probability on the other, with the actual number of rules in the transition table of the smallest Turing machine found to produce the string of interest the first time around in a well-specified enumeration of Turing machines from smaller to larger (see Fig.~\ref{instructions}).

CTM is able to suggest that some strings have lower algorithmic complexity than assigned them by lossless compression or LZW, relative to the highest complexity values for the set of strings in question, and to pinpoint the specific strings, indicating that it has the potential to spot objects likely to actually have lower algorithmic complexity that could not have been found with LZW or Shannon entropy. See Fig.~\ref{program} and Fig.~\ref{causalgaps}).

\subsection{Agreement of distributions for high frequency elements}

The result of CTM as a method is the execution of trillions of small Turing machines each producing an output, or being discarded if found not to halt by analytical means (e.g. the Busy Beaver~\cite{rado}), or to have taken longer than a runtime threshold arrived at by an educated guess. This execution produces a distribution based upon how many times each string produced has already been produced by other Turing machines, in increasingly larger rule spaces (defined either by number of states multiplied by symbols, usually binary, or simply by the number of transition rules used after execution). The result is what theoretically is known as the Universal Distribution, with each machine count producing, for each string, a number potentially representing an estimation of its algorithmic probability (i.e. the likelihood of finding a Turing machine that produces a string whose digits are each chosen with equal probability).

The fact that descriptions (computational models) turn out to be less relevant than expected after CTM when producing these empirical distributions that explore increasingly larger sets of Turing machines, or using models other than Turing machines, or simply different language specifications (see Fig.~\ref{models} and Fig.~\ref{instructions}), was not known and came as a surprise, as Chaitin pointed out in~\cite{report}:

\begin{quote}
    The theory of algorithmic complexity is of course now widely accepted, but was initially rejected by many because of the fact that algorithmic complexity depends on the choice of universal Turing machine and short binary sequences cannot be usefully discussed from the perspective of algorithmic complexity. [However, of CTM] $\ldots$  discovered, employing [t]his empirical, experimental approach, the fact that most reasonable choices of formalisms for describing short sequences of bits give consistent measures of algorithmic complexity! So the dreaded theoretical hole in the foundations of algorithmic complexity turns out, in practice, not to be as serious as was previously assumed.
\end{quote}

Indeed, results from CTM could not have been predicted without actually performing the massive experiments (some using the largest world super computers), to some extent because top frequency convergence in output distributions turned out to happen more often than theoretically expected, in the face of changes to the underlying universal reference (even if perhaps not optimal) computational model (see Fig.~\ref{models}).

The apparent stability of a distribution that could be affected by what is perhaps a multiplicative factor but instead looks like a logarithmic factor even smaller than the additive one, at least for the most frequent and purportedly least algorithmically complex strings as they keep emerging from longer rule spaces, is an open theoretical question, related to optimality and the invariance theorem. Indeed, what this seems to suggest is that the density of computational models (reference machines) that quickly converge in distributions dominates the density of those that are purposely designed to fail, and thus that the choice of model may not play such an important role in practice as the theory would have it when a `natural' choice is made (as opposed to a `constructed' reference machine). Such a conclusion could not have been reached otherwise, which shows how experimental science feeds back into theory and reflects the work around CTM and BDM. This epistemological point was originally suggested in~\cite{thesis}. Pushing boundaries at the edge of what is possible by mechanical means serves as a moving target both for experimentalists to explore the `incomputability frontier' and for theoreticians to relate their work to frontline challenges in the application of the theory to other areas of science.

The empirical distributions from CTM are not only relatively stable, but behave both in unexpected and expected ways. Many tests have been designed, conceived and successfully passed~\cite{kolmo2d,physicaa,computability,bdm}. So when dealing with short strings, statistical lossless compression is unsuitable, and when dealing with long strings it is irrelevant when linked to algorithmic complexity (beyond Shannon entropy), which is to say it under-performs in both (all) cases.

What research around CTM demonstrated is that apparently disparate computational models produce similar output distributions for high frequency elements (hence with large algorithmic probability) despite the possible theoretical counter-indications (see e.g. Fig.~\ref{models}). This is one of the advantages of approaching algorithmic complexity from a quasi-empirical standpoint, as attested to by Chaitin himself when evaluating CTM~\cite{report}.

For the first time, CTM offered access to empirical output distributions calculated from large sets of computer programs that could be tested against theoretical predictions~\cite{calude}. In diverse areas, CTM has found applications to network characterisation and causation where state-space description methods such as CTM are most relevant as they are able to provide computable models~\cite{abrahao,gauvrit,diogo}. 

\begin{figure}[ht!]
\centering
\scalebox{.48}{\includegraphics{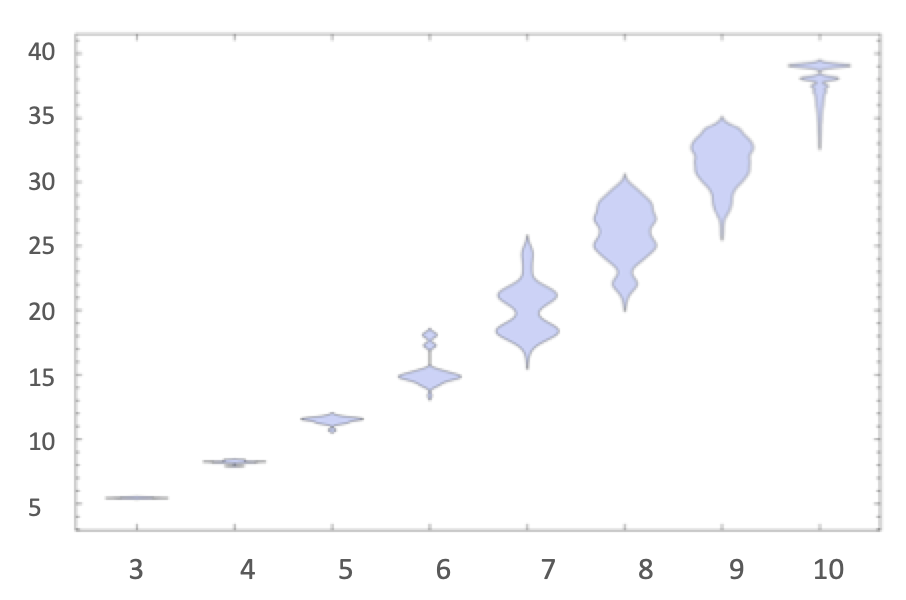}}
\caption{\label{droplike}Each `drop-like' distribution represents the set of strings that are minimally produced with the same number of instructions ($Y$ axis) for all machines with up to 4 states for which there can be up to $2(n+1)=10$ transition rules. The more instructions needed to produce the strings, the more complex they are shown to be according to CTM ($Y$ axis), after applying a coding-theorem like equation to the empirical probability values for each string. The results conform with the theoretical expectation-- the greater CTM, the greater the number of instructions used by the shortest Turing machine according to the enumeration, and vice-versa. Adapted from~\cite{computability}.}
\end{figure}

The agreement in complexity values between estimations by CTM and lossless compression (LZW) over a set of strings of growing length is very high. The interesting cases are, of course, those discrepant ones where entropy (and LZW) may diverge from CTM, that is, strings that display high entropy and low compressibility but are flagged by CTM as having shorter programs (shorter relative to the size of the rest of the programs in the same string set), as reported in~\cite{bdm} and in~\cite{nmi}.

\begin{figure}[ht!]
\centering
\scalebox{.45}{\includegraphics{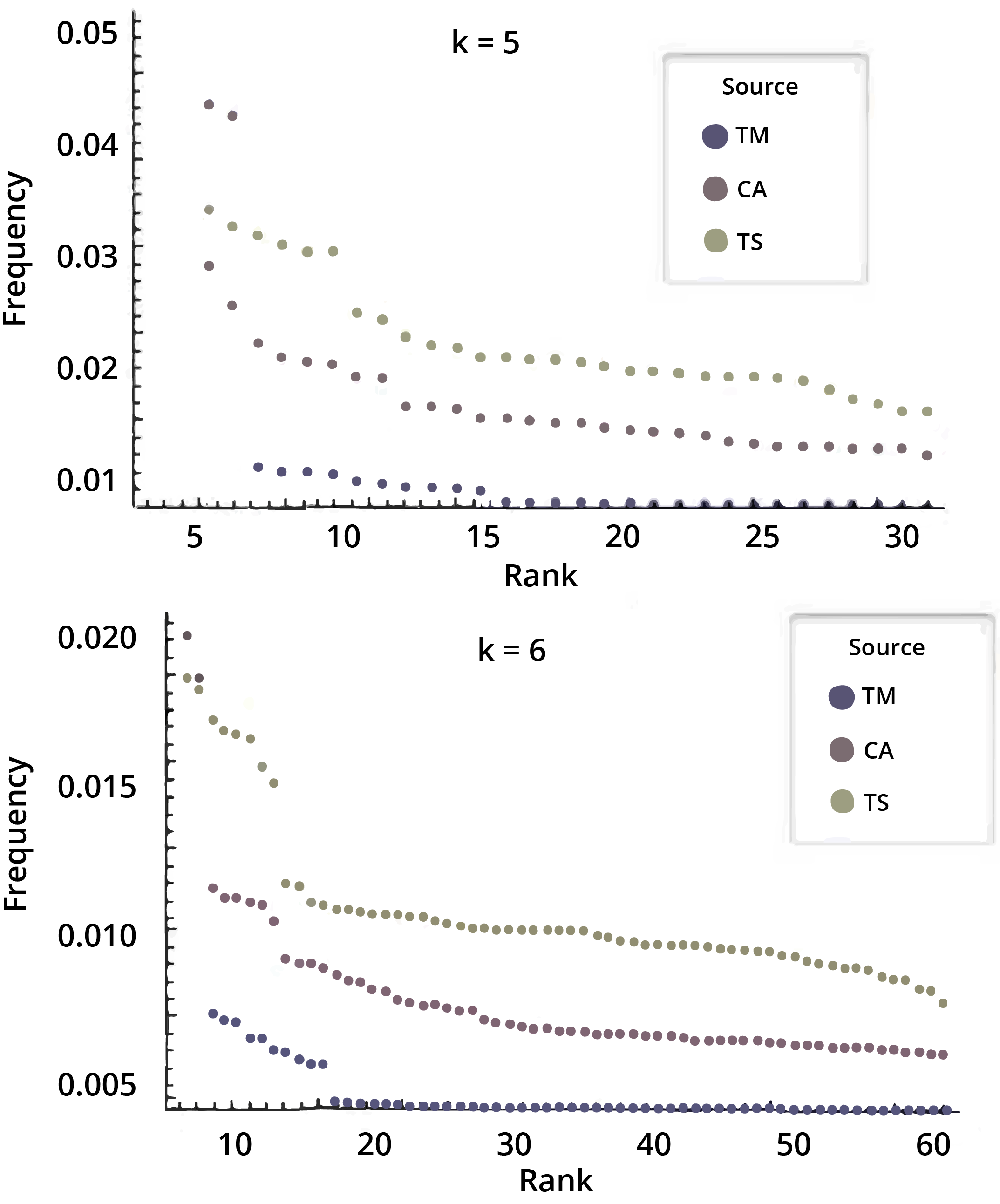}}
\caption{Agreement in shape and ranking of empirical distributions produced by 3 completely different models of computation: Turing machines (TM), Cellular Automata (CA) and Post Tag Systems (TS), all 3 Turing universal models of computation for the same set of strings of length $k=5$ and $k=6$ for purposes of illustration only (comparison was made across all string lengths up to around 20, which already includes a large set of $2^{20}$ strings).}
\label{models}
\end{figure}

\begin{figure}[ht!]
\centering
\scalebox{.5}{\includegraphics{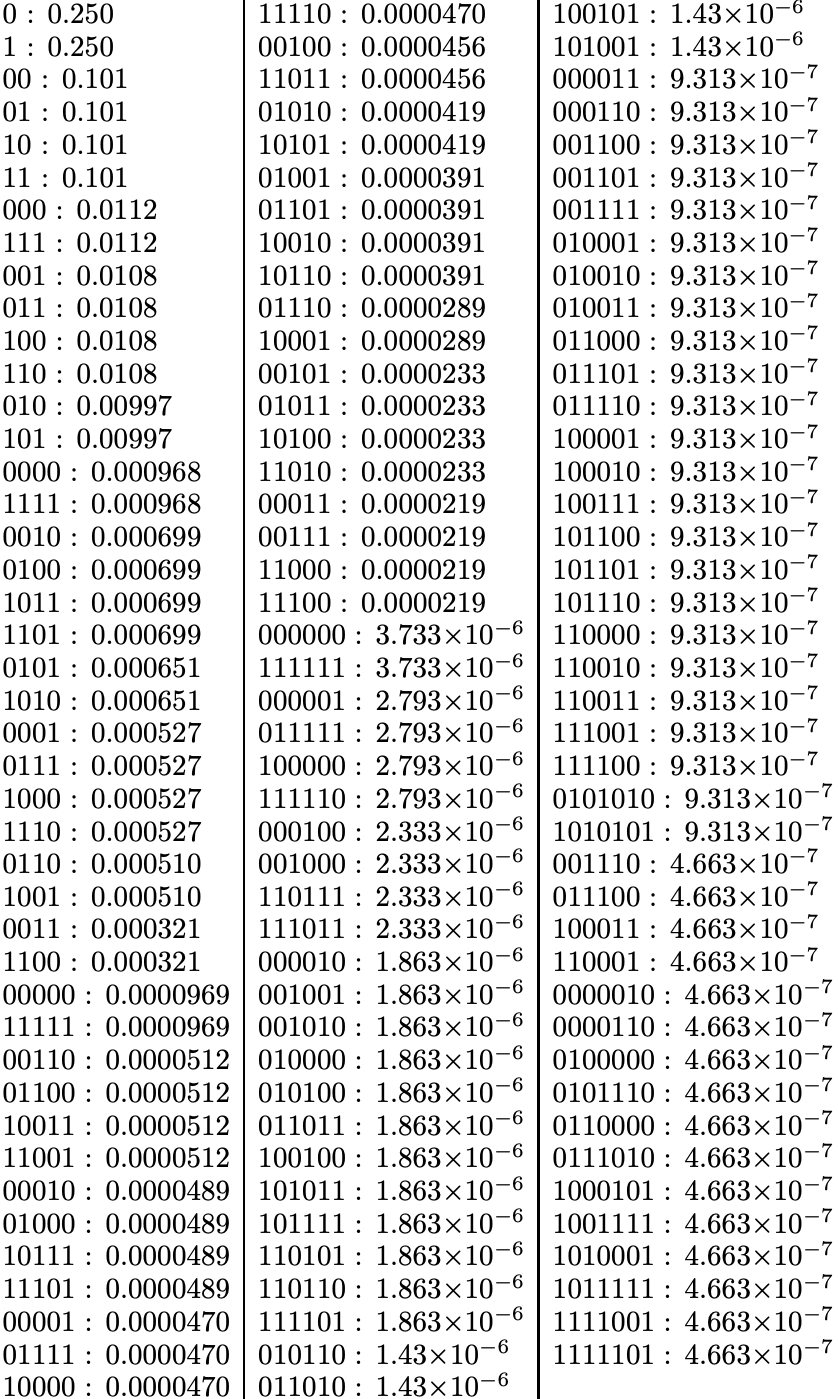}}
\caption{\label{empirical}This is the first ever empirical probability distribution produced by all the 15\,059\,072 Turing machines with 3 states and 2 symbols as reported in~\cite{d4}. CTM has produced several comprehensive large tables for all binary (and also non-binary~\cite{ploscomp}) Turing machines with up to 4 and 5 states (and currently for 6 states), and is currently also computing for 6 states on one of the largest super computers available. The distributions (this and the much larger ones that followed) have shed light on many matters}, from long standing challenges (the short string regime left uncovered by statistical compression), to the impact of change of choice of computational power on empirical output distributions, to applications tested in the field that require high sensitivity (such as perturbation analysis~\cite{iscience,nmi}).
\end{figure}

As shown in Fig.~\ref{empirical}, the output distribution from small Turing machines tends to classify strings by length (not even this relatively simple phenomenon could have been anticipated without actually running the experiment), with exponentially decreasing estimated algorithmic probability values. The distribution comes sorted by length blocks from which one cannot easily tell whether those at the bottom are more random than those in the middle, but one can definitely say that the ones at the top, both for the entire distribution and by length block, are intuitively the simplest. Both $0^k$ and its reversed $1^k$ for $n \leq 8$ are always at the top of each block, with 0 and 1 surmounting them all. There is a single exception in which strings were not sorted by length. This is the string group 0101010 and 1010101 found four places ahead of their length block, which we take as an indication of a complexity classification becoming more visible, since these two strings are what one would intuitively consider less random, because they are easily described as the repetition of two bits, as captured by Shannon entropy or compression.

\begin{figure}[ht!]
\centering
\scalebox{.46}{\includegraphics{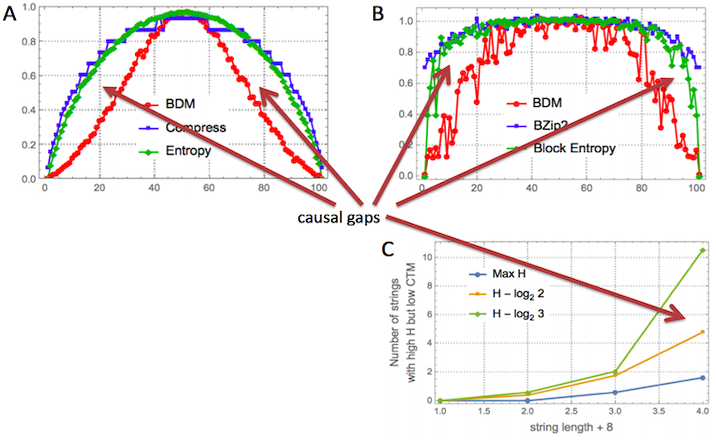}}
\caption{\label{causalgaps}A: Unlike compression, which behaves just like Shannon entropy, CTM approximates the long-tail shape of the Universal Distribution, better conforming with the theory under strong assumptions of optimality. We call these causal gaps (A and B) because they are precisely the cases in which strings can be explained by a computer program of length far removed from the greatest program length of the most algorithmically random strings according to our (universal) computational model. C: Divergence of CTM from entropy (and therefore LZW). Adapted from~\cite{bdm}.}
\end{figure}

\begin{figure}[ht!]
\centering
\scalebox{.27}{\includegraphics{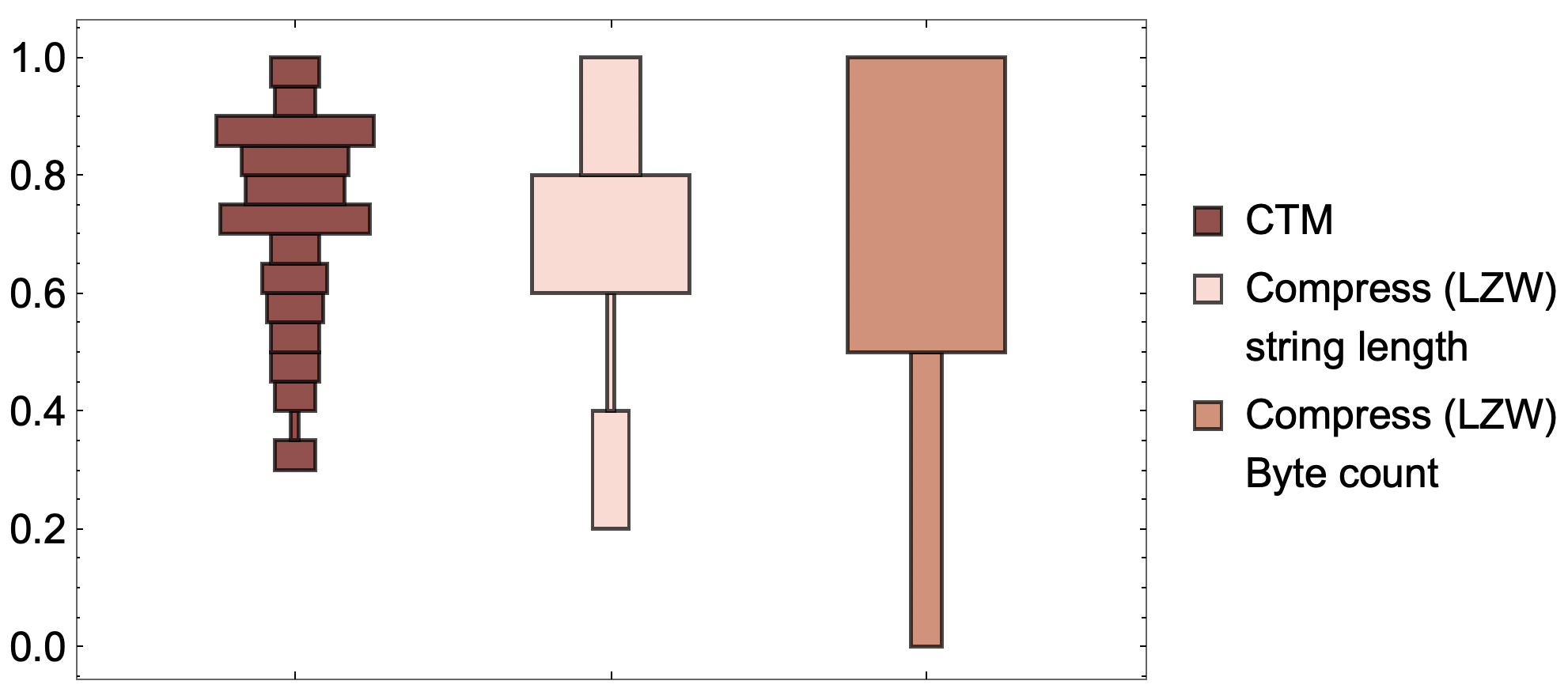}}
\caption{\label{diversity}Density histograms showing how poorly lossless compression performs on short strings (just 12 bits in length), collapsing all values into 2 to 4 clusters, while CTM produces a finer-grained classification of 36 groups of strings. CTM (and BDM) can be calculated using the Online Algorithmic Complexity Calculator (\url{https://complexitycalculator.com/}).}
\end{figure}

Fig.~\ref{causalgaps} shows how compression behaves like Shannon entropy and how CTM approximates algorithmic complexity, with results conforming with the theoretical expectation, and Fig.~\ref{diversity} shows how lossless compression algorithms collapse most values.

One may also ask how robust the complexity values and classifications may be in the face of greater changes in the computational model formalism or reference Universal Turing machine (e.g. Turing machines with several tapes, and all possible variations, or other universal means of computation). We have shown~\cite{algonat,kolmo2d} that radical changes to the computing model produce very similar distributions and ranks (see Fig.~\ref{models}), and we have even explored what happens with subuniversal systems at different levels of the Chomsky hierarchy~\cite{liliana} ranking distributions of complexity values (using completely different computing models). The models of computation in Fig.~\ref{models} include one with no natural halting state (CA). The agreement is greater among higher frequency strings, something that may be intuitive but is theoretically unexpected, given machine universality questions and the additive or multiplicative constants involved. Adapted from~\cite{algonat}.

Furthermore, we have proved that one of our methods that combines some of the proposals that we have advanced to approximate algorithmic complexity by way of algorithmic probability, called BDM, behaves like Shannon entropy in the worst case, while in the best case it does incorporate local estimations of algorithmic complexity (provided by CTM) and binds them together using classical information theory. In other words, we cannot do much worse than LZW because a version of LZW (block entropy) is considered our worst performing approach, given that, as we have  demonstrated, we cannot under-perform (by much) Shannon entropy at the task of finding only repetitions~\cite{bdm}.

\subsection{Problematic use of statistical arguments}

Papers on CTM (see e.g.~\cite{plos} and the sketch in Fig. 1 in ~\cite{bdm} for purposes of illustration) always open by explaining that the invariance theorem too has little relevance for applications (beyond its positive contribution of having given birth to algorithmic complexity by virtue of guaranteeing convergence at the limit). This is because it does not tell us anything about convergence rates, therefore making it impossible to guarantee whether we are approximating algorithmic complexity quickly or rather diverging before converging. Furthermore, the same short review~\cite{review} goes on to justify decades of the use of statistical lossless compression: `However, we assume [$\dots$] that the natural data we are dealing with contain no complicated mathematical constructs like $\pi= 3.1415\dots$ or Universal Turing machines'. 

By `natural data' the author presumably means `data from the natural or real world'. What's referred to as `constructs' in~\cite{review} are mathematical objects that the author seems to disdain or dismiss as impossible to encounter in the real world. However, many of these objects are widely regarded as ubiquitous in science and engineering, often associated with natural properties or physical processes. Furthermore, these are exactly the objects whose possible characterisation using algorithmic complexity distinguishes algorithmic information from classical (Shannon) information. The use and defence of statistical compression is convenient because it avoids having to deal with all the main challenges in the application of algorithmic complexity, namely incomputability, optimality, and the additive constant. Approaches that assume either that algorithmically compressed objects and /or Turing machines are constructs gives up on Turing-universality altogether, and in turn on algorithmic complexity itself, given that it is the very element in the theory of algorithmic information that distinguishes it from Shannon entropy (and from classical statistical compression).

This minireview~\cite{review} seems to defend a position that assumes that there are only statistical regularities in natural data~\cite{review} to justify the use of statistical compression algorithms in the last 2 decades, and that recursive functions such as $f(x):=x+1$ which can produce $s=123456\ldots$ (when initial value is $x=1$) with no statistical regularities if not transformed by another algoritm (it has been proven to be Borel normal~\cite{champernowne}) are artificial or unrealistic examples impossible to be found in nature or with no interest for humans to be properly characterised (e.g. as not being random). In the theory of algorithmic information, however, objects like $s$ or $\pi$ or the so-called Thue-Morse sequence that are recursively generated but do not produce trivial statistical patterns (like most computable numbers/processes with short descriptions but no trivial statistical redundancy), are quintessential instances of what algorithmic complexity was designed to tell apart: statistical (apparent) from algorithmic (non-apparent) randommness. If it were to be argued in favour of this stance that nature is full of noise, that does not seem to work either. If anything, it makes matters worse, because noise will tend to destroy any possible statistical pattern that some researchers would like to identify with a statistical compression algorithm, whereas the bulk of the algorithmic content can still be captured by a short function separating noise from compressed data. 
The minireview in~\cite{review} seems to suggest that any function $f$ that produces a string $s$ with the length of the description of $f$ much shorter than $s$ is a `construct' (as used in~\cite{review} to negatively denote an `artificial' or `unreal' nature) if it does not give away its nature by exhibiting some statistical redundancy. If applied to the 332 thousand integer sequences found (as of March 7, 2020) in the Online Encyclopedia of Integer Sequences (OEIS), Shannon entropy, LZW and other popular statistical lossless compression algorithms would fail at characterising the non-random nature of these highly recursive sequences because most of them do not display any repetitive segments. According to the viewpoint adopted in~\cite{review}, all these sequences would be labelled `constructs' and not objects that may be associated with natural or physical phenomena (from which, by the way, many of them derive, e.g. simple progressions of population growth or epidemic spread). In contrast, CTM and measures based on and motivated by the principles of algorithmic complexity can characterise properties of these sequences~\cite{integersequences}.

In fact, what the results from CTM suggest~\cite{d4,plos}, as has been fruitfully investigated, is that what appear to be constructs are ad hoc non-optimal universal machines, and that the choice of reference machine appears to be much less important than potentially expected from the theory, if it is not made non-optimal on purpose, as in designing a Turing machine conceived to assign strings specific complexity values with artificial additive or multiplicative terms like the one proposed recently in the review~\cite{review}. The chosen models produce similar output distributions despite the possible theoretical expectation (see, e.g., Fig.~\ref{models}). And this is one of the values of approaching algorithmic complexity from a quasi-empirical standpoint, as attested to by Gregory Chaitin himself when officially evaluating CTM~\cite{report}. Calude found empirical evidence from CTM that could provide pointers to new results on halting probabilities~\cite{calude2} as well as confirm and reexamine previous theoretical results. For the first time, access to an empirical output distribution (and ranking!) of small strings from a large set of computer programs was possible (see~\ref{empirical}), despite the challenges and the assumptions made (See Table~\ref{relax}). Today, dozens of papers resonate with this approach, in full awareness of the assumptions, but finding applications to network characterisation and causation and even inspiring new methodological algorithms~\cite{abrahao,gauvrit,diogo}. In cognition, CTM has been found to be an appropriate framework for approaching questions relevant to neuroscience. For example, it is known that humans can recall long sequences by virtue of reconstructing them from short, recursively generated descriptions such as $123456\ldots$, despite the latter not having any statistical regularities that Shannon entropy, LZW or any other statistical measure can characterise. An use of CTM in this domain can be found in~\cite{gauvrit}, singled out simply for purposes of illustration, as CTM has now been used in dozens of papers in the area of cognition. In all these papers it has always been openly disclosed (e.g.~\cite{plos} and Fig. 1 in ~\cite{bdm}) that the invariance theorem (which encompasses the problem of optimality) is of no consequence for applications. This is because the invariance theorem does not tell us anything about convergence rates, even under the assumption of optimality.

The review in~\cite{review} enters unexpected territory in the following passage: `However, we assume [$\dots$] that the natural data we are dealing with contain no complicated mathematical constructs like $\pi= 3.1415\dots$ or Universal Turing machines’ (presumably `natural data' means data from the natural world). The author refers to as a `construct'~\cite{review} an object widely regarded to be ubiquitous in science and engineering, associated with natural properties and physical processes. So instead of dealing with incomputability, optimality, and the additive constant, the approach that assumes that algorithmically compressed objects are constructs drops computational universality altogether, the most important ingredient in the theory of information and the feature that distinguishes that theory from say, Shannon entropy (and ultimately, statistical compression). The very theory underlying the field of algorithmic complexity, computability theory, also establishes that most objects are of the type that Vit\'anyi calls `constructs'~\cite{review}, and are the only objects in respect to which algorithmic complexity performs differently than, say, the most trivial statistical counting function such as Shannon entropy. These arguments, again, end up uncoupling statistical compression-based approaches from algorithmic complexity, associating them only with Shannon entropy. The review~\cite{review} closes the passage cited above by defending heuristic approaches based on Shannon entropy instead of the principles of algorithmic complexity: `In fact, we assume that the natural data we are dealing with contains primarily effective regularities that a good compressor finds.' In transiting to the next sentence it takes a huge leap: `Under those assumptions the Kolmogorov complexity of the object is not much smaller than the length of the compressed version of the object.' One of the approaches to algorithmic complexity apparently defended in~\cite{review} under the `natural data' argument is to give up on algorithmic complexity itself by relaxing the Turing universality requirement and not worrying about anything beyond statistical properties. Small wonder then that there is never any need to be concerned with any of the other dependency levels (from 0 to 4 as shown in Table~\ref{relax}), in particular, machine optimality.

In cognition and psychometrics, for example, CTM has been found to be the appropriate framework for asking questions relative to features other than trivial statistical ones~\cite{ploscomp}. Indeed, it is known that humans can recall long sequences by virtue of reconstructing them from short, recursively generated descriptions such as $123456\ldots$, despite it not having any statistical regularities that Shannon entropy, LZW, or any other statistical measure is able to characterise. An example of the use of CTM in this domain can be found in~\cite{gauvrit}, singled out simply for purposes of illustration, as CTM has now been used in dozens of papers in the area of cognition.

\subsection{Reality check: from theory to practice}

In~\cite{vitanyibook}, whose purpose is to find applications of algorithmic (Kolmogorov) complexity, a purported application to behavioural sequences is introduced to quantify a controversial experiment in the area of ant communication introduced by Reznikova in the 60s~\cite{reznikova}. However, the application of lossless compression algorithms fails because the strings are too short, making it necessary to resort to the reader's intuition, just as Reznikova herself did. The purpose was to demonstrate that communication time and path complexity in a maze were correlated: the more complicated the path the longer ant foragers would take to transmit instructions to other ants. In a paper using CTM~\cite{zenilmarshall}, however, it was demonstrated that it can be informative and practical to take on such a task involving short sequences and to say something meaningful and objective about them that conforms to intuitive and theoretical expectations, rather than having to appeal directly to the reader's intuition.

Some criticism~\cite{review} seems to stem from the fact that we adapted Levin's Universal Distribution (UD) and Solomonoff's algorithmic probability (AP) to yield a measure based on or motivated by UD and AP, while not using UD or AP themselves. This comes as a surprise, given that the author of the review~\cite{review} has himself been using LZW and other lossless compression algorithms based on Shannon entropy for about 3 decades. While the review~\cite{review} seems to take definitions very seriously when it comes to the coding theorem, it is only too flexible when it comes to the use of heuristics based on statistical compression algorithms. The review~\cite{review} constructs an argument that asks us to be more rigorous in the choice of algorithms to approximate algorithmic complexity while allowing much greater liberties to be taken and broader assumptions to be made when adopting statistical lossless compression, assuming all patterns are statistical (repetitions) or that Turing universality is irrelevant for applications of algorithmic complexity altogether. 

When moving from theory to practice, one has to compromise on aspects of a theory that are not feasible. This is the case even with general purpose computers, which differ from Turing machines with unbounded tape. Which doesn't mean that these aspects of the theory have been overlooked or are never meant to be improved upon as we make progress, leaving behind weaker approaches. 

In all papers related to CTM and BDM, it has always been maintained that how robust our methods are is an important question---that is, how sensitive they are to changes in the reference universal Turing machine or choice of incomputable enumeration---if we are to understand and possibly estimate the overhead added by the additive constant from the invariance theorem. We know that the invariance theorem guarantees that the values converge in the long term, but we have always acknowledged that the invariance theorem tells us nothing about rates of convergence.

Pending a closer inspection of the details of their proposal, the potential challenge of Bauwens, Mahklin, Vereshchagin and Zimand's approach~\cite{bauwens1} is its top-down nature. They propose to start from a target string and run a process to find a short list of short computer programs (in the given language) by cleverly filtering out programs that do not follow the specific defined graph path towards constructing the target string during runtime, thereby reducing the computational time required for the exploration to a small polynomial. Judging by the fact that a later publication by one of the leading authors recommended in the review~\cite{review} as one of the methods to look after settled on using LZW in a recent work~\cite{bauwens2}, it potentially means that the use and application of the recommended approach may be more difficult than expected. Before actual applications, not even these authors may be able to anticipate the ways in which their approach will have to compromise when it comes to real-world applications, nor should an untested technique be endorsed over CTM that has already confronted the challenge of making itself numerically relevant an offers the only alternative, for good and bad, to statistical compression. Indeed, it would seem that the application of this (and other) novel approaches are more difficult than expected, and they still need to find implementations. What appears exciting in the approach of Bauwens et al., the polynomial complexity of the process of finding short lists of short programs, will most likely turn out to be less tractable in practice than predicted in theory when dealing with real-world applications. This is because this bottom-up approach requires a search for each target string, for every case and for every application before computing the said short list of short computer programs--unless one adopts an approach similar to CTM, which involves turning the tables and precomputing a large set of short computer programs producing a large set of short strings combined by an approach that may need to be similar to BDM, a divide-and-conquer approach consisting of breaking down an original larger string into smaller substrings for which the algorithm is more tractable in practice. How this will pan out in reality remains to be seen. Whatever the case, it would appear that some ideas of CTM and BDM may, in the end, help implement others. The original method may still be infeasible for most purposes and practical applications but the final result may end up not being that different from the choices we had to make when we came up with CTM~\cite{d4} and BDM~\cite{bdm}, an approach that may be an upper bound less tight than the estimation directly on the longer string, but one that can be computed in a timewise practicable fashion and incrementally improved upon (by eventually increasing the computing time). The final value of the combination of CTM and BDM will be a sum arrived at by an educated guess (as opposed to a random value or one arrived at using only statistical compression algorithms). 

There are also other relevant and promising approaches to measures motivated by algorithmic complexity that should be explored and supported by the community as alternatives to statistical compression. Some of these include Solovay functions~\cite{bienvenue,bienvenue2}, minimal computer languages~\cite{brainfuck} and, approaches similar to ours, ~\cite{liliana}--model proposals involving restricted levels of computational power (but beyond statistical compression algorithms) such as Finite-state and transducer complexity~\cite{caludefsa,caludefsa2,caludetransducer}. We ourselves have produced a variant of a transducer complexity model in a CTM fashion~\cite{liliana}, while also introducing
other CTM approaches based on different models and computational power~\cite{liliana}.

In practice, to assign temporary algorithmic probabilities has no impact for the purposes for which CTM was designed, is used and can be empirically tested, as the measure will either fail or not in the domain of application (just as happens when statistical compression algorithms are used). Just as with any measure that is semi-computable, both the principles of the implemented measure and the methodology can incrementally be improved upon, and represent an upgrade from the particular limitations of statistical compression algorithms. The only other concern in~\cite{review} about CTM is the widely known additive constant that has dogged all methods and applications, and is a general concern, not one exclusive to CTM. Indeed it should be more of a concern in approaches like LZW, the operative question being: how will other lossless compression algorithms impact the values? let alone how they will miss all the algorithmic non-statistical properties of data for which indeed a shorter description would be sufficient test for non-randomness and by itself a short computer program but most likely that shorter program will not be found whatsoever if the data did not display statistical features (redundancy) in the first place (which is something Shannon entropy alone would have been able to do in the first place). For CTM, it has been shown that radical changes to the underlying model of the reference Universal Turing machine produce very similar distributions or have little effect on the strings ranked highest by algorithmic probability~\cite{algonat,plos}. Yet, it is dishonest to operate using statistical lossless compression algorithms with total impunity without proper disclaimers and, more important, false pretences on the basis of alleged connections to algorithmic complexity which are only circumstantial. Statistical lossless compression algorithms not only face the same challenges, they are not even computationally universal are thus not up to the task or properly fit for purpose. The additive constant is simply maximised by the many arbitrary decisions that must be made when designing a statistical compression algorithm whose original purpose was unrelated with implementing a method for approximating algorithmic complexity, being related to algorithmic complexity only to the degree that Shannon entropy is. Indeed, LZW has been proven to be `universal'~\cite{lzw} in the sense that for an arbitrary window length such an algorithm would reach the Shannon entropy rate of any object, but it is not `universal' in the computability sense as basic requirement to instantiate algorithmic complexity. 

One can also see how LZW has been adjusted in practice, as it is practically impossible to implement an arbitrary sliding window length to traverse a string looking for repetitions, forcing one to settle on a maximum fixed length, since computers, at the end of the day, have finite memory. So the remarks in~\cite{review} regarding these two concerns would be equivalent to alleging that LZW is wrong (even for the purpose it was designed for) simply because it is unable to implement an arbitrary sliding window length which is a compromise of its own instantiation.

\subsection{Limitations and challenges of all approaches}

Users seeking applications of algorithmic complexity are faced with a dearth of choices: either they continue the (mis)use and abuse of statistical compression algorithms in the context of algorithmic complexity in their application domain, or undertake methodological exploration of other measures such as CTM and BDM and use them when appropriate, since it has been proved that at their worst these can only behave as badly as a version of Shannon entropy (block Shannon entropy, similar to LZW)~\cite{bdm}. CTM is no stranger to compromise or to questions about optimality and the constant from the invariance theorem. These are indeed challenges that we have openly acknowledged (e.g.~\cite{plos} when disclosing the assumption that the invariance theorem holds).

However, the use of black-box approaches such as statistical compression algorithms does not circumvent any of these challenges, but rather obfuscates them. These are approaches that cannot be defended except on weak grounds, such as that they avoid Turing universality.
They amount to settling for an option not (much) better than (or a trivial variation of) Shannon entropy. In contrast, CTM and BDM have the advantage of embracing Turing universality and the problems that come with it, of being based on or motivated by algorithmic principles beyond statistics, and the ability to find short computer programs (shortest under the chosen model), thus being closer in all respects and for all purposes to the principles of algorithmic complexity.

Without the use of black boxes such as lengths of compressed binary files that have no state space correspondence, the agreement of ranking distributions of estimations of complexity values of strings produced by CTM and compressed with algorithms such as LZW to strings of increasing length is almost perfect, with statistical rank correlation values close to 1 (thereby conforming with the theory, as we know that low entropy also means low algorithmic complexity). The interesting cases are, however, those in which compression fails (the short string regime) and those where strings are found to have high entropy but low algorithmic complexity (also conforming with the theory, as we know that high entropy does not imply low algorithmic complexity). So CTM and BDM behave as theoretically and intuitively expected~\cite{plos,bdm}.

The review~\cite{review} points out what its author believes are drawbacks of CTM, but these are in fact the very features that have made them applicable and which have been widely tested, thus making for a real alternative or complement to statistical compression algorithms (see Table~\ref{regimes2}). Among the claims made in~\cite{review} is that the sum of the empirical probability values produced by CTM is not smaller than 1, as the theoretical version of the Universal Distribution definition would require. This was part of the original design, in order to be able to assign fully determined probability values for practical purposes as a computable version~\cite{integersequences} of the incomputable measure on which it is based  ( while making no pretense to being identical to it). It was decided to construct a dynamic lookup table that can be incrementally updated every time a new short computer program is found under the specific reference Universal Turing machine model (in our case, a non-computable enumeration sorted by TM state $\times$ symbol size). But the computable process itself generates probability values whose sum is 1, and thus when running a subsequent round by expanding the rule-space of Turing machines these probabilities are updated (each probability value remaining equal or becoming smaller than its equivalent in the previous round). Taken separately, however, every empirical distribution would add to the overall probability, making it larger than 1. The remarks only confirm the fact that CTM, motivated by AP, does what it was designed to do, without pretending to be AP itself. However, every time that the pre-computation for CTM takes place, those probability values are constantly updated and in the limit they would all sum not more than 1.

On the one hand, while lossless compression explores the most radical of the approaches by relaxing the most basic of the necessary conditions for algorithmic complexity to be different from classical information theory, that of the Turing-universality requirement thereby collapsing all features in data to statistical features and equating both algorithmic and classical information theories (both in theory and practice), what CTM does is to embrace Turing universality and take on relaxing properties lower in the necessary conditions-hierarchy (e.g. those marked with numbers 1 and 2 on Table~\ref{relax}). On the other hand, we do know what the consequences of embracing statistical compression like LZW are, and that only assuming that  simple recursive processes producing successions like $1,2,3,4,5,6,\ldots$ (or the Thue-Morse sequence and most computable sequences) are `constructs' of too artificial nature impossible to appear in human or natural data. In this scenario, statistical compression algorithms do not provide a greater power on previously used and less onerous methods such as simple cognates of Shannon entropy. Statistical lossless compression does not have to concern itself with properties and challenges of the algorithmic dependency-hierarchy (Table~\ref{relax}), instead remaining outside its realm not even dealing with the first level of complications.

\begin{table}[!htb]
\begin{center}
\begin{tabular}{c|c|l|l}
	\textbf{Dependency}&\textbf{Relaxed}&\textbf{Challenge/}&\textbf{Positive }\\
	\textbf{level}&\textbf{property}&\textbf{limitation}&\textbf{implication}\\
	\hline
	0&Universality & some form of  & characterisation\\
	& & incomputability & beyond statistical\\
	\hline
	1&Optimality & ad hoc models & ranking invariant\\
	& & & in the limit\\
	\hline
	2&Invariance & no rate of  & invariant  in the\\
	& & convergence & limit w/overhead\\
	\hline
	3&Prefix-freeness & ad hoc language & slightly tighter\\
	& &  & bounds\\
	\hline
	4&Coding theorem & 0 to 3 have & (AP for C and C \\
	& & to hold & for AP) $\pm$ O(1)\\
\end{tabular}
\label{relax}
\end{center}
\caption{List of nested theoretical properties in increasing rank of necessary condition to instantiate algorithmic complexity (lower rows depend on the properties of higher rows) and their potential negative and positive implications when holding. Different measures explore the implications of the relaxation of each of these properties. Proofs of optimality are nontrivial~\cite{hardness} if the model is not ad hoc, designed for the purpose of separating data from model (as MML does), because the model is simple (statistical) and avoids Turing universality). Rates of convergence of additive constants are never guaranteed, and any approach will suffer from it. Statistical lossless compression algorithms give up on universality at the very first level and therefore can claim to avoid all others. CTM embraces 0 (thus the first is the most important), assumes 1 and thus 2; 3 is of minor concern~\cite{review} and thus 4 can be applied. Statistical lossless compression such as LZW, however, can be seen as embracing triviality as it simply does not deal with even level 0 and can only capture data redundancy in the form of repetitions.}
\end{table}

\subsection{Relaxing necessary conditions and studying the implications}

While the problem of the incomputability of algorithmic complexity, the additive constant and the non-conformance between practice and theory are not avoided by using statistical lossless compression algorithms, as their use may suggest, the limitations of computable approaches to semi-computable functions have never been concealed in the context of CTM and BDM. In~\cite{plos}, for example, the most standard specification of a reference universal Turing machine was used and properly defined. The heading of the Online Algorithmic Complexity Calculator (OACC) reads, in part, as follows:

\begin{quote}
    $\ldots$ also, very important, the Numerical Limitations subsection in the \textit{How It Works} subpage.
\end{quote}

And in the above mentioned subsection on Limitations it is stated that:

\begin{quote}
    Numerical limitations of CTM are the ultimate incomputability of the universal distribution, and the constant involved in the invariance theorem which, nevertheless, we have quantified and estimated to apparently be under control, with values converging even in the face of computational model variations. Our papers cover these limitations and their consequences should be taken into account.
\end{quote}

The online calculator also includes comparisons to other measures, including Shannon entropy and compression using gzip (LZW), for purposes of illustration.

When we establish the limitations of CTM and BDM, we also make the following disclosure:

\begin{quote}
    For BDM, the limitations are explored in~\cite{bdm}, and they are related to boundary conditions and to the limitations of CTM itself. The paper also shows that when CTM is not updated, BDM starts approximating Shannon entropy over the long range, yet the local estimations of CTM shed light on the algorithmic causal nature of even large objects.
\end{quote}

To address these valid concerns that are common in science when making assumptions, and indeed never go away, all sorts of sanity checks to verify at least model consistency were performed~\cite{computability,physicaa,kolmo2d,plos,bdm} (also see Fig.~\ref{instructions}). In Fig.~\ref{droplike}, CTM is shown to be in agreement with the minimum number of instructions used by the first Turing machine in the enumeration to produce the string for the first time, as investigated and reported in~\cite{computability}. Thus two methods to produce algorithmic complexity-based values are in high agreement.

The authors of the aforementioned review have explored the consequences of relaxing certain theoretical requirements, as we ourselves have done, not only in~\cite{d4,plos} but also in~\cite{liliana}. And ours are weaker assumptions than those made when using statistical compression algorithms, such as that (1) the type of data in nature is of a statistical nature only, excluding `constructs' like $\pi$ (so called in~\cite{review}), or (2) assumptions about the kind of machinery nature can or cannot instantiate (such as Turing universality). These kinds of relaxing condition approaches are common in science and mathematics and studying their consequences yields useful insights and, in the case of CTM, yield measures grounded or at least motivated by the principles of algorithmic information that do not assume that, e.g., only statistical features are to be found in natural data.

Regarding CTM, it was inquired whether expanding the rule space or exploring further along incomputable enumeration would affect the empirical probability distribution~\cite{plos}. Because of the lack of convergence rate from the invariance theorem The theoretical expectation could have indicated that changes would have a major effect, for example, even inverting the ranking of highly frequent elements (especially under conditions of non-optimality of the reference models), but this was not the case. We found that increasing the size of the rule space respects the original ranking, except for some value discrepancies and mostly only for least frequent strings (therefore the ones with highest randomness and most unstable ones, given the small number of machines generating these strings). This was not obvious and a completely different result could have been expected and produced despite the fact that all Turing machines with a lower number of states are included in all larger rule spaces (larger machines also includes those that do not use one of the states). Different rankings could have been produced because the number of machines in larger rule spaces exponentially exceeds the number of machines in the smaller rule space. However, the agreement between rule spaces is more stable than theoretically expected~\cite{plos,kolmo2d}, suggesting that the overhead is actually smaller and perhaps even reducible (one cannot ever know for sure because of the invariance theorem).

\subsection{Algorithmic complexity in scientific discovery}

In advancing CTM and BDM what was sought was to introduce a measure able to capture algorithmic features from data, beyond statistical features. It sought to contribute to model-driven data analysis. At the core of the research that CTM made possible was not only the goal of estimating algorithmic complexity for its own sake, which is no doubt interesting, but in the interests of the kind of inspection and analysis that one can perform by simulating computable processes to search for sets of computable models able to explain a piece of data assembled for causal analysis~\cite{iscience} see Fig.~\ref{program} and Fig.~\ref{causalgaps}). The analysis could then include an examination of the correspondence of program and domain states (computational v. physical) to help understand possible computable underlying mechanisms of causation explaining an object or the evolution of a dynamical system~\cite{iscience,nmi}. Such an approach also has implications for areas such as Machine Learning, which is currently heavily reliant on and motivated by statistical approaches very similar in spirit and just as problematic as the heuristics of compression algorithms. In contrast, what we did in a recent paper~\cite{nmi} was to explain how algorithmic complexity could help to deconvolve data beyond statistics by searching for computable models. In our programme, values of algorithmic complexity are only used as a guiding tool, not as a goal, but our greatest interest is in generating the set of testable computable models from a universal non-computable process to understand first principles underlying observed data, something that LZW, as a black-box approach, would be incapable of doing in any meaningful way, as it would not provide any computer program able to match any state variable of the domain under review (e.g. compressing planetary movements with LZW would never give you anything even close to a physical law), beyond the trivial quantification of the number of repetitions in the data.

We took on these challenges in order to advance beyond LZW, which does not embrace universality and assumes objects like $\pi$ to be constructs; and because we are less interested in the actual estimations of a single value for algorithmic complexity and more in the process of producing generative computable models to explain data whose study can shed light on the underlying state variable mechanisms, which in combination with perturbation analysis we call Algorithmic Information Dynamics~\cite{iscience}. In this context, CTM is very similar to~\cite{bauwens1}, in that we explore and produce a list of short programs for each target string.

In this sense, CTM and what it has led to, such as Algorithmic Information Dynamics~\cite{nmi,iscience,scholarpedia}, is similar to, the algorithmic response to, or the generalisation of areas that study state-space modelling like Dynamic Causal Modelling~\cite{DCM} or Computational Mechanics~\cite{crutchfield}, where the interest is in finding and analysing generative mechanisms as explanations of observed data. As such, possible estimations of algorithmic complexity are almost a byproduct of the exploration of the state space of computable hypotheses with CTM.

\section{Conclusions}

Access to complementary methods is key to covering the spectrum of different short and long string regimes necessary for wider applications (see Table~\ref{regimes2}). Indeed, when applied to short strings, LZW and other statistical compression algorithms would fail, so they are found to be of very limited use in the short string regime. For example, in an approach that tested and compared compression in a problem of molecular biology, it was shown that CTM and BDM can be informative of nucleosome occupancy~\cite{nar} when applied to short strings (a nucleosome of DNA is only around 146 base pairs long), a key feature in genetic regulation. The problem of nucleosome occupancy is one of the greatest challenges in molecular biology, believed to be second only to protein folding. Many applications can now be found for methods aimed at this short string regime.

\begin{table}[!htb]
\begin{center}
\begin{tabular}{c|c|c|c}
	\textbf{Method}&\textbf{Regime }&\textbf{Outcome}&\textbf{Capability/}\\
	\textbf{}&\textbf{of application}&\textbf{}&\textbf{reach}\\
	\hline
	LZW, MDL, MML & long  & length of an often &statistical only\\
    and similar& strings & obfuscated file&\\
	\hline
	CTM or similar &  & large set of small &algorithmic\\
     & short strings & computer programs & \\
    &  & generating the data & \\
	\hline
	CTM & short and  & none compared to & algorithmic \\
	$+$ BDM& long strings &  alternatives& and statistical\\
\end{tabular}
\end{center}
\caption{Different methods for different length regimes and for  different purposes. Here LZW represents the set of statistical lossless compression algorithms.}
\label{regimes2}
\end{table}

As even Cilibrasi has pointed out~\cite{personal}, CTM provides an approach to a long-standing problem in an area that lacked a methodology to cover the regime of short strings. Statistical methods capture redundancy in the form of string repetitions, with no connection to state variables, unlike the generation of a set of algorithmic/mechanistic generative mechanisms from CTM, that can be regarded as computable models of the data. CTM$+$BDM combines the best of the statistical and algorithmic approaches. Notice that algorithmic also implies statistical because any statistical feature is also (trivially) algorithmic, though the reverse is not true. However, CTM$+$BDM has the advantage of being able to find long statistical patterns that CTM alone would miss, while CTM has the potential to characterise the algorithmic content of short sequences that BDM alone (and the rest of the statistical models) would, left to their own devices, miss.

An open question is why output probability distributions generated by completely different reference universal Turing machines turn out very similar, despite concerns related to optimality and additive---or other (if not optimal)---constants which resonates with Chaitin's optimism when learning about CTM~\cite{report}. I call this the ``Unreasonable Effectiveness of Digital Computation in the Natural World''~\cite{chaitinfest}. As is the case in other areas of mathematics, it could be that we are too careful (as no doubt we should be), whereas in practice, things do not correspond to theoretical expectations. This should have been the case of, for example, G\"odel's undecidability theorems across all areas of mathematics; yet undecidability is rarely found or an impediment for doing mathematics. Most of the time, solutions are found to open problems within some axiomatic framework, thereby circumventing undecidability.

The question of optimality may therefore most likely be the real construct, judging by the robustness of the distributions generated by different models of computation, as quantified by us. Indeed, one hypothesis is that most `natural' universal machines will be optimal or semi-optimal and produce a `natural' order in output distribution. It remains to be seen whether these possible `natural' models chosen are inadvertently biased in some consequential way, making them produce similar output probability distributions, and whether this possible bias affects---and by how much---the output distributions that appear to converge, in particular, for the highest frequency strings, when in theory they simply should not. This may not only be interesting in its own right, but as a different, guided and quasi-empirical approach to asking and answering highly theoretical questions.

Paradoxically, the weakest approaches and the ones most unrelated to applications of algorithmic complexity are those based on statistical lossless compression algorithms that dominate the field, for the wrong reasons, inhibiting methodological innovations in a field that requires them. Recent theoretical results published in ~\cite{bauwens1} and ~\cite{teutsch} are examples of the kind of innovation needed in the field, and certainly do not disqualify others. On the contrary, they can both build upon other innovations. The manifest advantage of the approach of Bauwens et al.~\cite{bauwens1} that is covered almost exclusively in~\cite{review} (apart from  CTM and a brief but bold defence of statistical compression) is that it may better control the associated additive constant and may possibly circumvent the optimality requirement. 

One of the several points that the minireview in~\cite{review} missed to mention is that we are still currently faced with a mostly binary choice---judging by the limited coverage of any other alternatives in~\cite{review}---for picking a method to deal with algorithmic complexity: (1) those based on statistical lossless compression algorithms such as LZW that have been used (and abused) for decades and which the author of the minireview in~\cite{review} has himself exploited for at least two decades; or (2) CTM (augmented with BDM), which offers an alternative and has started to be used in more areas, appears closer to the principles of algorithmic information theory and has addressed a problem uncovered by statistical lossless compression, the regime of short strings.

In~\cite{review}, a lot of trust is placed in a potential alternative, yet it will be tested once it is made available for numerical application, and when it is, it  may face its own challenges. As for the statistical lossless compression approach, we have argued that it is too weak to instantiate the basic principles of algorithmic complexity.

Future efforts to measure algorithmic complexity should focus less on obtaining single real-value numbers at any cost, as when using statistical lossless compression, and instead work to imbue with a deeper purpose applications meant to help with fundamental questions in science. Such an approach does not preclude addressing theoretical challenges, but not at the expense of innovation, especially where available options are weaker and of less relevance to questions of great interest to science, such as causation.


\begin{thebibliography}{99}
\bibitem{franklin} Franklin, J.N.Y.;  Porter, C.P. Key developments in algorithmic randomness. {\it arXiv} {\bf 2004}, arXiv:2004.02851.

\bibitem{bienvenu3} Bienvenu, L.; Shafer, G.; Shen, A. On the history of martingales in the study of randomness. \textit{Electron. J. Hist. Probab. Stat.} {\bf 2009}, {\it 5}, 1.

\bibitem{kolmo}  Kolmogorov, A.N. Three approaches to the quantitative definition of information. \emph{Probl. Inf. Transm.} {\bf 1965}, {\it 1}, 1--7.

\bibitem{martin}  Martin-L\"of, P. The definition of random sequences. \textit{Inf. Control} {\bf 1966}, {\it 9}, 602--619.

\bibitem{davis} Davis, M. \textit{The Universal Computer, The Road from Leibniz to Turing.} W. Norton \& Company: New York, NY, USA, 2000.

\bibitem{caludebook}  Calude, C.S. \textit{Information and Randomness An Algorithmic Perspective, Texts in Theoretical Computer Science. An EATCS Series}; Springer-Verlag Berlin Heidelberg: Heidelberg, Germany, 2002.

\bibitem{nies} Nies,  A. \textit{Computability and Randomness}. Oxford University Press: Oxford, UK, 2009.

\bibitem{downey}  Downey, R.G.; Hirschfeldt, D.R. \textit{Algorithmic Randomness and Complexity, Theory and Applications of Computability}; Springer-Verlag: New York, NY, USA, 2010.

\bibitem{vitanyibook}
~Li, M.; Vit\'anyi, P.~
\newblock {\em An Introduction to Kolmogorov Complexity and Its Applications};
\newblock Springer: Heidelberg, Germany, 2008.

\bibitem{review}  Vit\'anyi, P.M.B. How incomputable is Kolmogorov complexity? \textit{Entropy} {\bf 2020}, {\it 22(4)}, 408.

\bibitem{lzw}  Ziv, J.;  Lempel, A. Compression of individual sequences via variable-rate coding. \textit{IEEE Trans. Inf. Theory}, {\bf 1978}, {\it24 (5)}, 530.

\bibitem{algebraic1}  Dongarra, J.J.; Du Croz, J.; Hammarling, S.; Hanson, R.J. A proposal for an extended set of Fortran Basic Linear Algebra Subprograms. \textit{ACM SIGNUM Newsl.} {\bf 1985}, {\it 20}, 2--18.

\bibitem{algebraic2} Ancis M.; Giusto, D.D. Image data compression by adaptive vector quantization of classified wavelet coefficients. \textit{IEEE Pac. Rim Conf. Commun. Comput. Signal Process. PACRIM} {\bf 1997}, {\it 1}, 330--333.

\bibitem{datacompression} Salomon, D. \textit{Data Compression: The Complete Reference}; Springer Science \& Business Media: Berlin/Heidelberg, Germany, 20 March 2007.

\bibitem{borel}  Borel, E. Les probabilit\'es d\'enombrables et leurs applications arithm\'etiques, \textit{Rendiconti del Circolo Matematico di Palermo} {\bf 1909}, {\it 27}, 247--271.

\bibitem{cilibrasi}  Cilibrasi, R.L.;  Vit\'anyi, P.M.B. Clustering by compression.  \textit{IEEE Trans. Inf. Theory} {\bf 2005}, {\it 51:4}, 1523--1545.

%\bibitem{langton} C.G. Langton, Studying artificial life with cellular automata, \emph{Physica D: Nonlinear Phenomena 22} (1--3): 120--149, 1986.

\bibitem{chaitin}
 Chaitin, G.J.
\newblock On the length of programs for computing finite binary sequences: Statistical considerations.
\newblock {\em J. ACM} {\bf 1969}, {\it 16}, 145--159.

\bibitem{zenilentropy} Zenil,  H. Towards Demystifying Shannon Entropy, Lossless Compression, and Approaches to Statistical Machine Learning. In Proceedings of  International Society for Information Studies (IS4IS) summit, University of California, Berkeley, CA, USA, 2--6 June 2019.

\bibitem{antunes}  Teixeira, A.; Matos, A.; Souto A.;  Antunes, L. Entropy Measures vs. Kolmogorov Complexity. \textit{Entropy} {\bf 2011}, {\it 13},  595--611.

\bibitem{solo}  Solomonoff, R.J. Complexity-Based Induction Systems: Comparisons and Convergence Theorems, \textit{IEEE Trans. Inf. Theory} {\bf 1978}, {\it 24}, 422--432.

\bibitem{solo2}  Solomonoff, R.J. The Application of Algorithmic Probability to Problems in Artificial Intelligence.  \textit{Mach. Intell. Pattern Recognit.} {\bf 1986}, {\it 4}, 473--491. 

\bibitem{solo3}  Solomonoff, R.J. A System for Incremental Learning Based on Algorithmic Probability. In Proceedings of the Sixth Israeli Conference on Artificial Intelligence, Computer Vision and Pattern Recognition, Tel Aviv, Israel, December 1989; pp. 515--527. 

\bibitem{levin}  Levin, L.A. Universal sequential search problems. \textit{Probl. Inf. Transm} {\bf 1973}, {\it 9(3)}, 265--266.

\bibitem{miracle}  Kirchherr, W.; Li, M.;  Vit{\'a}nyi, P.
\newblock The miraculous universal distribution.
\newblock {\em The Math. Intell.} {\bf 1997}, {\it 19(4)}, 7--15, 1997.

\bibitem{solovay} Solovay,  R.M. Draft of paper (or series of papers) on Chaitin’s work. Unpublished notes, 215 pages, May 1975. Also in  Downey, R.G.; Hirschfeldt, D.R. \textit{Algorithmic Randomness and Complexity, Theory and Applications of Computability}; Springer-Verlag: New York, NY, USA, 2010.

\bibitem{antunes2}  Antunes, L.; Fortnow, L. Time-Bounded Universal Distributions, Electronic Colloquium on Computational Complexity. Report No. 144, 2005.

\bibitem{minsky}  Minsky, M. Panel discussion on The Limits of Understanding. \textit{World Science Festival}, NYC, Dec 14, 2014. Available online: \url{https://www.worldsciencefestival.com/videos/the-limits-of-understanding/} (accessed on 26 February 2020). 

\bibitem{algonat}  Zenil, H.; Delahaye, J-P. On the Algorithmic Nature of the World. In \textit{Information and Computation}; Dodig-Crnkovic, G., Burgin, M., Eds.; World Scientific Publishing Company: Singapore, 2010.

\bibitem{d4} Delahaye, J.-P.; Zenil, H.
\newblock Numerical evaluation of algorithmic complexity for short strings: A glance into the innermost structure of randomness. \newblock {\em Appl. Math. Comput.} {\bf 2012}, {\it 219}, 63--77.

\bibitem{plos}  Soler-Toscano, F.; Zenil, H.;  Delahaye, J.-P.; Gauvrit, N.
Calculating Kolmogorov Complexity from the Output Frequency Distributions of Small Turing Machines
\textit{PLoS ONE 9(5)}: e96223, {\bf 2014}.

\bibitem{levinkt}  Levin, L.A. Randomness conservation inequalities; information and independence in mathematical theories, \textit{Inf. Control} {\bf 1984}. {\it 61}, 15--37.

\bibitem{buhrman}  Buhrman, H.; Fortnow, L.;  Laplante, S. Resource-Bounded Kolmogorov Complexity Revisited. \textit{SIAM J. Comput.} {\bf 2001}, {\it 31(3)}, 887--905.

\bibitem{allender} Allender,  E.; Buhrman, H.; Kouck\'y, M.; van Melkebeek, D.; Ronneburger, D. Power from random strings. \textit{SIAM J. Comput.} {\bf 2006}, {\it 35}, 1467--1493.

\bibitem{schmid2}  Schmidhuber, J. The Speed Prior: A New Simplicity Measure Yielding Near-Optimal Computable Predictions. In Proceedings of the International Conference on Computational Learning Theory COLT 2002: Computational Learning Theory, ; Kivinen, J., Sloan, R.H., Eds.; Springer: New York, NY, USA, 2002; pp. 216--228. Sydney, NSW, Australia, 8--10 July 2002; pp. 216--228.

\bibitem{hutter} Hutter, M. \newblock {\em Universal artificial intelligence: Sequential decisions based on algorithmic probability}. \newblock Springer Science \& Business Media: Berlin/Heidelberg, Germany, 2004.

\bibitem{wallace}  Wallace, C.S.; Boulton, D.M. An information measure for classification. \textit{Comput. J.} {\bf 1968}, {\it 11}, 185--194.

\bibitem{rissanen}  Rissanen, J. Modeling by shortest data description. \textit{Automatica} {\bf 1978}, {\it 14 (5)}, 465--658.

\bibitem{liliana}  Zenil, J.; Badillo, L.;  Hern\'andez-Orozco, S.; Hernandez-Quiroz, F. Coding-theorem Like Behaviour and Emergence of the Universal Distribution from Resource-bounded Algorithmic Probability. \textit{Int. J. Parallel Emerg. Distrib.. Syst.} {\bf 2018}, doi:10.1080/17445760.2018.1448932.

\bibitem{chomsky}  Chomsky, N. Three models for the description of language, \textit{IEEE Trans. Inf. Theory} {\bf 1956}, {\it (2)}, 113--124.

\bibitem{crutchfield}  Shalizi, C.R.;  Crutchfield, J.P. Computational mechanics: Pattern and prediction, structure and simplicity. \textit{J. Stat. Phys.} {\bf 2001}, {\it 104}, 817--879.

\bibitem{schnorr}  Schnorr, C.P. A unified approach to the definition of a random sequence. \textit{Math. Syst. Theory} {\bf 1971}, {\it 5 (3)}, 246--258.

\bibitem{schnorr2}  Schnorr, C.P. Process complexity and effective random tests. \textit{J. Comput. Syst. Sci.} {\bf 1973}, {\it 7 (4)},  376--388.

\bibitem{calude1} Calude, C.S.;  Longo, G. The deluge of spurious correlations in big data. {\it Found. Sci.} {\bf 2017}, {\it 22}, 595--612.

\bibitem{smalldata}  Zenil, H. Algorithmic Data Analytics, Small Data Matters and Correlation versus Causation. In \textit{Berechenbarkeit der Welt? Philosophie und Wissenschaft im Zeitalter von Big Data}; Ott, M., Pietsch, W., Wernecke, J., Eds.; Springer Verlag: New York, NY, USA, 2017; pp. 453--475.

\bibitem{rado}  Rad\'o, T. On non-computable functions. \textit{Bell Syst. Tech. J.} {\bf 1962}, {\it 41 (3)},  877--884.

\bibitem{bdm}  Zenil, H.;  Hern\'andez-Orozco, S.;  Kiani, N.A.;  Soler-Toscano, F.;  Rueda-Toicen, A. A Decomposition Method for Global Evaluation of Shannon Entropy and Local Estimations of Algorithmic Complexity, \textit{Entropy} {\bf 2018}, {\it 20(8)}, 605.

\bibitem{hardness} Bienvenu,  L.;  Desfontaines, D.;  Shen, A. Generic algorithms for halting problem and optimal machines revisited. {\it Log. Methods Comput. Sci.} {\bf 2015}, {\it  12}, 1--29.

\bibitem{integersequences} Soler-Toscano, F.; Zenil, H. A Computable Measure of Algorithmic Probability by Finite Approximations with an Application to Integer Sequences. \textit{Complexity} {\bf 2017}, {\it  2017}, 7208216.

\bibitem{nmi} Zenil, H.;  Kiani, N.A.; Zea, A.; Tegn\'er, J. Causal Deconvolution by Algorithmic Generative Models. \textit{Nat. Mach. Intell.} {\bf 2019}, {\it 1}, 58--66.

\bibitem{iscience} Zenil, H.; Kiani, N.A.; Marabita, F.; Deng, Y.; Elias, Y.;  Schmidt, A.; Ball, G.;  Tegn\'er, J. An Algorithmic Information Calculus for Causal Discovery and Reprogramming Systems. \textit{iScience} {\bf 2019}, {\it 19}, 1160--1172.

\bibitem{nar} Zenil, H.;  Minary, P. Training-free Measures Based on Algorithmic Probability Identify High Nucleosome Occupancy in DNA Sequences. \textit{Nucleic Acids Res.} {\bf 2019}, {\it 47}, e129.

\bibitem{bauwens1} Bauwens, B.;  Makhlin, A.; Vereshchagin, N.;  Zimand, M. Short lists with short programs in short time.  \textit{Conf. Comput. Complexity} {\bf 2018}, {\it 27}, 31--61.

\bibitem{scholarpedia}  Zenil, H.; Kiani, N.A. Algorithmic Information Dynamics, \textit{Scholarpedia}, 2019. Avaialble online: \url{http://www.scholarpedia.org/article/Algorithmic_Information_Dynamics}. (accessed on 20 March 2020)

\bibitem{DCM}  Friston, K.J.; Harrison, L.; Penny, W. Dynamic causal modelling. \textit{NeuroImage} {\bf 2003}, {\it 19 (4)}, 1273--1302.

\bibitem{champernowne}  Champernowne, D.G. The construction of decimals normal in the scale of ten. \textit{J. Lond. Math. Soc.} {\bf 1933}, {\it 8 (4)},  254--260.

\bibitem{report} Chaitin,  G. Evaluation report on the PhD thesis submitted Hector Zenil to the University of Lille ''Une approche exp\`erimentale \`a la th\'eorie de la complexit\'e algorithmique'' to obtain the degree of Doctor in Computer Science, 5/25/2011. Available online: \url{http://www.mathrix.org/zenil/report.pdf} (26 february 2020)

\bibitem{thesis} Zenil, H.  thesis to obtain the PhD in Computer Science, \textit{Une approche exp\`erimentale \`a la th\'eorie de la complexit\'e algorithmique}, University of Lille 1, France, July 2011.

\bibitem{calude2}  Calude, C.S.;  Dumitrescu, M. A probabilistic
anytime algorithm for the halting problem. \textit{Computability } {\bf 2018}, {\it7}, 259--271.

\bibitem{abrahao}  Abrah\~ao, F.S.; Wehmuth, K.; Ziviani, A. Algorithmic Networks: Central time to trigger expected emergent open-endedness, \textit{Theor. Comput. Sci.} {\bf 2019}, {\it  785},  83--116.

\bibitem{gauvrit} Mathy, F.; Fartoukh, M.; Gauvrit, N.; Guida, A. Developmental abilities to form chunks in immediate memory and its non-relationship to span development. \textit{Front. Psychol.} {\bf 2016}, {\it 7}, 201. 

\bibitem{diogo}  Silva, J.M.; Pinho, E.;  Matos, S.;  Pratas, D. Statistical Complexity Analysis of Turing Machine tapes with Fixed Algorithmic Complexity Using the Best-Order Markov Model. {\it Entropy} {\it 2020}, {\it 22 (1)}, 105.

\bibitem{kolmo2d}  Zenil, H.; Soler-Toscano, F.; Delahaye, J.-P.;  Gauvrit, N.  Two-dimensional Kolmogorov complexity and an empirical validation of the Coding Theorem Method by compressibility. {\it PeerJ Comput. Sci.} {\bf 2015}, {\it 1}, e23.

\bibitem{physicaa}  Zenil, H.; Soler-Toscano, F.; Dingle, K.; Louis, A. \newblock Correlation of automorphism group size and topological properties
 with program-size complexity evaluations of graphs and complex networks.
\newblock {\em Phys. A Stat. Mech. Its Appl.} {\bf 2014}, {\it 404}, 341--358.

\bibitem{computability} Soler-Toscano,  F.; Zenil, H.; Delahaye, J.-P.;  Gauvrit, N. Correspondence and Independence of Numerical Evaluations of Algorithmic Information Measures. \textit{Computability} {\bf 2013}, {\it 2}, 125--140. 

\bibitem{ploscomp}  Gauvrit, N.;  Zenil, H.; Soler-Toscano, F.;  Delahaye, J.-P.;  Brugger, P. Human Behavioral Complexity Peaks at Age 25. \textit{PLoS Comput. Biol.} {\bf 2017}, {\it13(4)},  e1005408. 

\bibitem{calude} C.S. Calude,
 Stay, M.A. Most programs stop quickly or never halt. \textit{Adv. Appl. Math.} {\bf 2007}, {\it 40}, 295--308.

\bibitem{reznikova}   Ryabko, B.; Reznikova, Z. \textit{Using Shannon Entropy and Kolmogorov Complexity to study the communicative system and cognitive capacities in ants Complexity}; John Wiley \& Sons Inc.: New York, NY, USA, 1996; Volume 2, N 2; pp. 37-42.

\bibitem{zenilmarshall}  Zenil, H.; Marshall, J.A.R.; Tegnér, J. Approximations of Algorithmic and Structural Complexity Validate Cognitive-behavioural Experimental Results. In {\it Alternative Computing}; Adamatzky,  A., Ed.; World Scientific: Singapore, 2020 

\bibitem{bauwens2}  Filatov, G.;  Bauwens, B.;  Kert\'esz-Farkas, A. LZW-Kernel: Fast kernel utilizing variable length code blocks from LZW compressors for protein sequence classification. \textit{Bioinformatics} {\bf 2018}, {\it 34(19)}, 3281--3288.

\bibitem{bienvenue}  Bienvenu, L.; Downey, R.; Nies, A.; Merkle, W. Solovay functions and their applications in algorithmic randomness. \textit{J. Comput. and Syst. Sci.} {\bf 2015}, {\it 81(8)}, 1575--1591.

\bibitem{bienvenue2} Bienvenu, L.;  Downey, R.; Nies, A.; Merkle, W. Solovay functions and K-triviality, 28th International Symposium on Theoretical Aspects of Computer Science (STACS 2011), pp 452--463. In T. Schwentick and C. D{\"u}rr, Leibniz International Proceedings in Informatics (LIPIcs), vol. 9, 2011. Available online: \url{https://hal.inria.fr/hal-00573598/} (26 february 2020)

\bibitem{brainfuck}  Cibej, U.;  Robic, B.; Mihelic, J. Empirical estimation of the halting probabilities. In Proceedings of the Computability in Europe (Language, Life, Limits), Budapest, Hungary, 23--27 June 2014.

\bibitem{caludefsa}  Calude, C.S.; Salomaa, K.;  Roblot, T.K. Finite state complexity. \textit{Theor. Comput. Sci.} {\bf 2011}, {\bf 412},  5668--5677.

\bibitem{caludefsa2}  Calude, C.S.;  Salomaa, K.;  Roblot, T.K. State-size Hierarchy for Finite-state Complexity. \textit{Int. J. Found. Comput. Sci.} {\bf 2012}, {\it 23}, 37--50.

\bibitem{caludetransducer}  Calude, C.S.; Salomaa, K.;  Roblot, T. Finite-State Complexity and the Size of Transducers. In Proceedings DCFS 2010, \textit{EPTCS} 31, 2010, pp. 38--47, 2010, doi:10.4204/EPTCS.31.6.

\bibitem{personal}  Cilibrasi, R. personal communication, March 7, 2020.

\bibitem{chaitinfest}  Zenil, H. Compression is Comprehension, and the Unreasonable Effectiveness of Digital Computation in the Natural World. In {\it Unravelling Complexity (Gregory Chaitin's 70 festschrift)};  Wuppuluri, S., Doria, F., Eds.; World Scientific Publishing: Singapore, 2019.

\bibitem{teutsch}  Teutsch, J. Short lists for shortest descriptions in short time, \textit{Comput. Complex.} {\bf 2014}, {\it 23},  565--583.

%\bibitem{santiago} S. Hern\'andez-Orozco, H. Zenil, J. Riedel, A. Uccello, N.A. Kiani, J. Tegn\'er, Algorithmic Probability-guided Supervised Machine Learning on Non-differentiable Spaces, arXiv:1910.02758 [cs.LG], 2019.
\end{thebibliography}
\end{document}